\documentclass[twocolumn]{aastex631}

\usepackage{multirow}

\usepackage{soul}

\begin{document}

\title{Magnetoacoustic Shocks and Spectropolarimetric Signals in He I 10830 {\AA}}
\author[0000-0001-9631-1230]{Hirdesh Kumar}
\author[0000-0003-1732-6632]{Tobias Felipe}
\author[0000-0002-3242-1497]{Christoph Kuckein}
\author[0000-0002-6546-5955]{S. J. González Manrique}
\author[0000-0002-1248-0553]{A. Asensio Ramos}

\affiliation{Instituto de Astrofísica de Canarias, 38205 C/ Vía Láctea, s/n, La Laguna, Tenerife, Spain}
\email{hkumar@iac.es}

\affiliation{Departamento de Astrofísica, Universidad de La Laguna, 38205 La Laguna, Tenerife, Spain}







\begin{abstract}

Umbral flashes are manifestations of magnetoacoustic shocks in the solar
chromosphere. These phenomena are thought to influence the evolution of
chromospheric umbral magnetic fields. However, the impact of these shocks on
inferred chromospheric magnetic field oscillations remains unclear. We examined
five different sunspots located near the solar disk center, observed with the GRIS
instrument installed at the GREGOR telescope. The \textsc{HAZEL2} Spectropolarimetric 
inversion code is used to obtain the  photospheric and chromospheric line-of-sight 
velocities and magnetic fields in Si 10827 {\AA} and He 10830 {\AA} spectral lines, 
respectively, using various inversion strategies. In the inversions with one chromospheric component, three of the sunspots exhibit remarkably stronger magnetic fields accompanying the shocks, while the other two sunspots show striking reductions in the magnetic field. Alternatively, the Stokes profiles can be reproduced by models with two chromospheric slabs, one on top of the other, through two-component inversions. These inversions provide excellent fits even when magnetic field fluctuations are discarded by imposing a constant magnetic field during the whole temporal series. In this scenario, the observed Stokes profiles are interpreted as the result of strong velocity gradients, where the He 10830 {\AA} line is sensitive to both sides of the shock front. Both competing models explaining the spectral profiles during the shocks, either large magnetic field fluctuations or velocity gradients, are critically discussed.

\end{abstract}



\section{Introduction} \label{sec:intro}

Magnetohydrodynamic (MHD) waves play an important role in the dynamics and heating of the solar atmosphere. In homogeneous plasma, these MHD modes are classified as slow and fast magnetoacoustic waves and the Alfvén mode \citep{Goedbloed}. In the inhomogeneous solar atmosphere, this terminology is still commonly used. Various characteristics of these modes have been reported and examined in detail in the atmospheres of different magnetized structures such as magnetic networks, plages, pores, and sunspots, using velocity and intensity observations from both space- and ground-based telescopes \citep{2006ApJ...640.1153C, 2010ApJ...722..131F, 2011ApJ...735...65F, 2016ApJ...830L..17Z, 2019A&A...621A..43F,
2019ApJ...871..155R, 2023A&A...676A..77F, 2023JASTP.24706071K, 2015LRSP...12....6K, 2015ApJ...812L..15K, 2017ApJ...842...59J, 2021RSPTA.37900172G, 2022ApJ...927..201A}. Upward-propagating slow magnetoacoustic waves dissipate their energy in the higher solar atmosphere in the form of shocks \citep{2006ApJ...640.1153C}. In some chromospheric lines, this appears as a sudden brightening, popularly known as umbral flashes. \cite{1969SoPh....7..351B} and \cite{1969SoPh....7..366W} reported the first detection of such phenomena in observations of a sunspot umbra in the Ca II K and H lines. The dominant period of umbral flashes has been found to be around three minutes. With advancements in high-resolution observations and non-local thermodynamic equilibrium (NLTE) inversion codes, different physical properties of umbral flashes have been revealed \citep{2000ApJ...544.1141S, 2000Sci...288.1396S, 2009ApJ...696.1683S, 2013A&A...556A.115D, 2015ApJ...800..129M, 2018A&A...614A..73F, 2018A&A...619A..63J, 2019ApJ...882..161A, 2019A&A...632A..75F, 2020A&A...642A.215H, 2020ApJ...892...49H, 2020ApJ...896..150Y, 2021A&A...645L..12F, 2023ApJ...945L..27F, 2025A&A...693A.165F}. Closely related to umbral flashes, running penumbral waves are also observed as concentric wavefronts propagating radially outward in the penumbra. Both of these phenomena are believed to be associated with upward propagating magnetoacoustic waves \citep{2007ApJ...671.1005B}. 

Additionally, magnetoacoustic waves have been reported to affect the chromospheric magnetic fields in sunspots in many works based on observations in the Ca II 8542 {\AA} spectral line and NLTE inversions with the code NICOLE \citep{2015A&A...577A...7S}. \cite{2013A&A...556A.115D} found chromospheric magnetic field fluctuations of the order of 200 G in the penumbra, which were associated with the passage of running penumbral waves. However, they did not detect significant fluctuations in the umbral magnetic field. \cite{2017ApJ...845..102H} found that during umbral flashes magnetic field is reduced compared to the quiescent phase. They suggested that the increase in the gas pressure due to the shock pushes the magnetic field lines away from the flash region. \cite{2018A&A...619A..63J} found fluctuations in magnetic field of the order of 270 G and 100 G in umbral flash and running penumbral regions, respectively. However, the magnetic field inferences from NLTE inversions of the Ca II 8542 {\AA} line have been challenged by \cite{2021ApJ...918...47F}. Utilizing MANCHA3D simulations \citep{2006ApJ...653..739K,2010ApJ...719..357F,2024SoPh..299...23M} of wave propagation in an umbral atmosphere and NLTE spectral synthesis and inversions, they found that the degeneracy of the Ca II 8542 {\AA} inversions can produce misleading magnetic field variations. This highlights the need for careful interpretation while analyzing magnetic field oscillations inferred from this spectral line. Chromospheric magnetic field oscillations in sunspots have also been explored using the He I 10830 {\AA} triplet. \cite{2018ApJ...860...28H} detected oscillations in the transverse component of the magnetic field during umbral flashes. Also, \cite{2023A&A...676A..77F} observed striking fluctuations in the chromospheric line-of-sight magnetic field. They interpret them as the result of changes in the response height of the He I 10830 {\AA} line to the magnetic field.\\
	
In this article, we present a comprehensive investigation of the chromospheric umbral magnetic field fluctuations during the magnetoacoustic shocks inferred from the spectropolarimetric analysis of the photospheric Si I 10827 {\AA} and the chromospheric He I 10830 {\AA} spectral lines in five different sunspots. We aim to confirm or refute the strong magnetic field fluctuations observed by \cite{2023A&A...676A..77F} and critically evaluate their interpretation as fluctuations in the atmospheric layers probed by the He I 10830 {\AA} triplet. The article is organized as follows: Section \ref{Sect2} addresses the observations and analysis, whereas in Section \ref{Sect3} we present the results. The discussion and conclusions are presented in the Sections \ref{Sect4} and \ref{Sect5}, respectively.

\section{Observations and data analysis}
\label{Sect2}

\subsection{GREGOR observations}

\begin{figure}
	\centering
	
\includegraphics[width=0.55\linewidth]{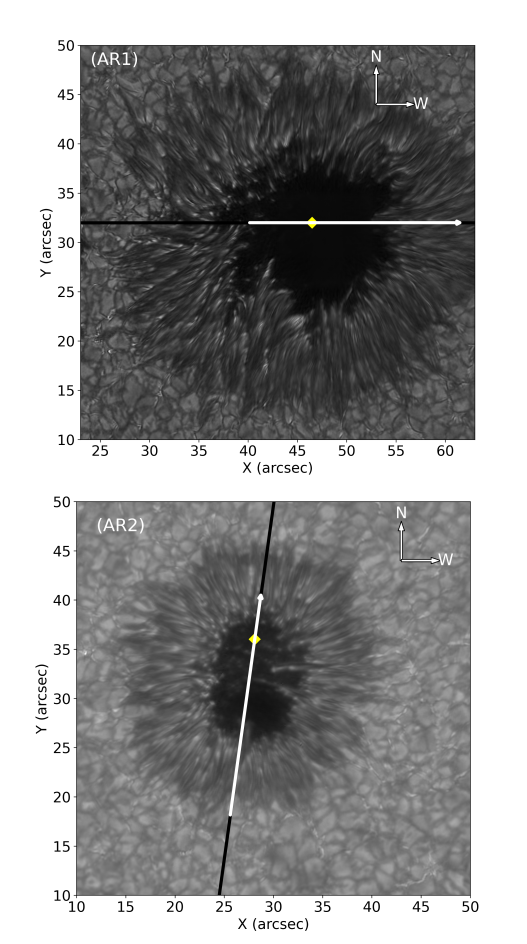}
\\[0.1cm]	
\includegraphics[width=0.55\linewidth]{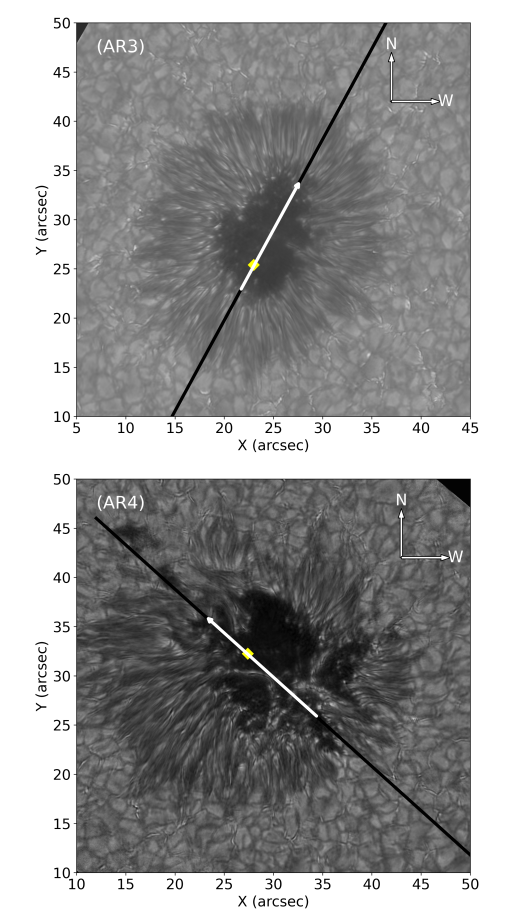}	
\\[0.1cm]
\includegraphics[width=0.55\linewidth]{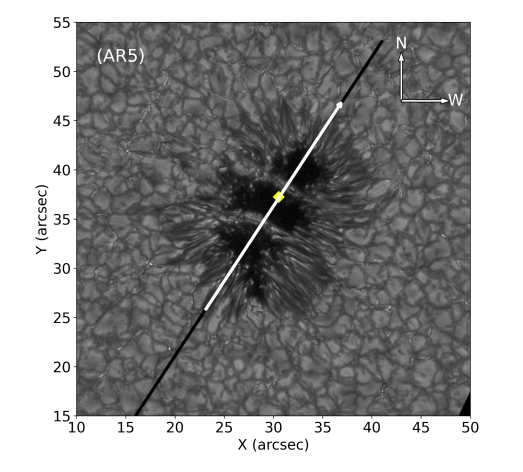}
			
	\caption{Sample G-Band images of the photosphere obtained from the HiFI and HiFI+ instruments showing the approximate position of the GRIS slit (black and white line) for all active regions. The white regions of the lines represent the inverted areas, as illustrated in Figure \ref{AR12345_Bz_Vlos_Map}, in which the arrows are oriented such that their heads indicate the upward direction in Figure \ref{AR12345_Bz_Vlos_Map}. The diamonds denote the pixel used to illustrate the temporal evolution of the velocity and magnetic field in Figure \ref{AR12345_Vlos_Bz_Temp}.}
	\label{AR12345_HiFi}%
\end{figure}

We have selected five different sunspots located near the disk center of the Sun to examine the magnetic field fluctuations in the chromospheric umbral regions during shocks. Table  \ref{exampleARtable} shows the list of active regions observed from the GREGOR Infrared Spectrograph (GRIS: \citealt{2012AN....333..872C}) attached to the GREGOR telescope \citep{2012AN....333..796S, 2020A&A...641A..27K}. We note that AR2 and AR3 are the same active region, but observed on subsequent days. In all cases, the slit was located at a fixed position crossing the sunspot during the whole temporal series. Figure \ref{AR12345_HiFi}  shows the sample G-band images of all active regions, obtained from the High-resolution Fast Imager (HiFI: \citealt{2018ApJS..236....5D}) and Improved  High-resolution Fast Imager (HiFI$+$: \citealt{2023JATIS...9a5001D}) instruments installed at the GREGOR telescope. The raw HiFI data consisted of fast bursts of 500 images, which were reduced using the sTools pipeline \citep{2017IAUS..327...20K}. Then, the best 100 frames of each burst where restored using the speckle-interferometry code KISIP \citep{2008SPIE.7019E..1EW}, producing one single restored image. The approximate position of the slit is shown with the black line. The full Stokes parameters ($I$, $Q$, $U$, and $V$) were recorded by GRIS along the whole slit (pixel size of $0.^{\prime\prime}135$), with the spectral range including the photospheric Si I 10827 {\AA} and chromospheric He I 10830 {\AA} lines. The raw GRIS data were corrected for dark current, flat fields, polarimetric calibration, wavelength calibration, and correction of residual fringes following the standard data reduction procedures \citep{1999ASPC..184....3C, 2003SPIE.4843...55C}. 

\begin{table*}
	\caption{Details of the active regions observed from the GREGOR Telescope and analysed in this study.}
	\label{exampleARtable}
	\centering
	\begin{tabular}{llllll}
		\hline
		\textbf{Labels} & \textbf{AR NOAA} & \textbf{$\mu = \cos\theta$}  & \textbf{Date} & \textbf{Cadence (s)} & \textbf{Time duration (UT)}  \\
		\hline
		 AR1 & 13679  &   0.90  & 2024 May 16 & 8.0 & 07:55:47 - 08:23:50  \\
		 AR2 & 12662  &   0.96  & 2017 June 18 & 5.6 & 08:02:00  - 09:36:00 \\
		 AR3 & 12662  &     0.89   & 2017 June 17 & 5.7 & 08:05:21  - 09:22:04 \\
		 AR4 & 13674  &  0.97   & 2024 May 17 & 8.1 & 08:00:56  - 08:26:44 \\
		AR5 & 14090 &    0.98 & 2025 May 21 &  6.3 & 07:45:13  - 08:55:41 \\
		\hline
	\end{tabular}
	
\end{table*}

\subsection{Spectropolarimetric inversions}

The information on photospheric and chromospheric line-of-sight velocities and magnetic fields comes from the Si I and He I spectral lines, respectively. These are inferred using the Hanle and ZEeman Light v2.0 (\textsc{HAZEL2}: \citealt{2008ApJ...683..542A}) Spectropolarimetric inversion code. For the modeling of the He I triplet, the code uses a constant property slab model. This slab is located at a height 'h' above the solar surface. The model assumes that the slab is uniformly illuminated from below by the photospheric continuum radiation, and the code takes into account atomic level polarization, and both the Zeeman and Hanle effects. The slab properties are described by several parameters: the line-of-sight velocity (V$_{LOS}$), three components of the magnetic field (B$_{x}$, B$_{y}$, B$_{z}$), optical depth ($\tau$) measured in the red component of the multiplet, Doppler width ($\Delta$v), enhancement factor ($\beta$), and damping parameter ($a$). The Helium’s triplet energy levels are treated with multiple fine-structure terms, rather than a simplified two-level approximation, using the so-called multi-term approximation \citep{2004ASSL..307.....L}. The telluric line at around 10832 {\AA} is modeled using a Voigt function in those data set (AR1--AR4) where it is included in the spectral range.  \\
	
\textsc{HAZEL2} can synthesize and invert multiple lines simultaneously. It handles the photospheric Si I line under the local thermodynamic equilibrium (LTE) assumption using the Stokes Inversion based on Response functions (SIR: \citealt{1992ApJ...398..375R}) code. In the inversion mode, SIR iteratively modifies the stratification of an initial guess atmosphere at selected locations in optical depth (so-called nodes). The hot11 model from \cite{1994A&A...291..622C} is used as the initial atmospheric guess for the photosphere. We employed three cycles, where the output of each cycle in employed as initial guess atmosphere for the following cycle. Table \ref{inversion_scheme} shows the distribution of nodes for the different physical parameters for the inversion of the Si 10827 {\AA} line, whereas in the case of the He I triplet, since \textsc{HAZEL2} uses a constant-slab model, inverted parameters are labeled as 1 and non-inverted parameters as 0.

We have also included stray light in our inversion, which is estimated from the spatial and
temporal averaging over a quiet region surrounding the sunspot. The inversion
assumes that the radiation detected at every pixel is an outcome of the contribution from the actual radiation coming from that spatial location and the signal from the spurious light, i.e., stray light. The code characterizes the amount of stray light with a filling factor. We have chosen to set a filling factor dependent on the spatial location along the slit but constant in time. In order to determine the value of the filling factor, we have first performed an inversion where the filling factor is a free variable. The median value of the filling factor over time at every pixel along the slit is then imposed to a second inversion of the profiles at that spatial location. This is done to minimize the impact of stray light variations on the inferred magnetic filed fluctuations. In the following, we will discuss the results of this second inversion.

In this work, we explore two different approaches for our inversion scheme: (i) 1-component inversions, where a single chromospheric slab is employed and (ii) 2-component inversions, in which we employed two chromospheric slabs, one on top of the other. In the latter, the radiation leaving from chromosphere one (ch1) is the incident radiation for chromosphere two (ch2). The 2-component inversion scheme has been applied to some selected profiles. In these cases, the chromospheric magnetic field has been fixed to the median value at the same spatial position estimated from the outcome of the 1-component inversions, while variations in the rest of the inverted parameters were allowed.

\begin{table}
	\caption{Number of nodes selected for the inversion of photospheric Si 10827 {\AA} and chromospheric He 10830 {\AA} spectral lines.}
	\label{inversion_scheme}
	\centering
	\begin{tabular}{llll}
		\hline
		\hline
		\textbf{Physical Parameter} &  Cycle 1 & Cycle 2 & Cycle 3 \\
		\hline
		\hline
		\textit{Photospheric Si 10827 {\AA} line} & & & \\
		\hline 
		
		Temperature & 3 &3 &3 \\
	   V$_{LOS}$ & 1 &2 &4 \\
		B$_{x}$ & 1 &2 &3 \\
		B$_{y}$ & 1 &2 &3 \\
		B$_{z}$ & 1 &2 &3 \\
		V$_{mic}$ &1 &2 &2 \\
		V$_{mac}$ &1& 0&0 \\
		
		\hline
		\textit{Chromospheric He 10830 {\AA} line} & & & \\
		\hline 
		
		 V$_{LOS}$ & 1 &0 &0 \\
		B$_{x}$ & 0 &1 &1 \\
		B$_{y}$ & 0 &1 &1 \\
		B$_{z}$ & 1 &1 &1 \\
		$\tau$ &1 &0 &0 \\
		$\Delta$v &1& 0&0 \\
		a &1& 0&0 \\
		\hline
		
	\end{tabular}
\end{table}

\section{Results}
\label{Sect3}

\subsection{Evolution of chromospheric line-of-sight velocity and magnetic field in 1-component inversions}

\begin{figure*}[h!]
	\centering
	
		\includegraphics[width=0.99\linewidth]{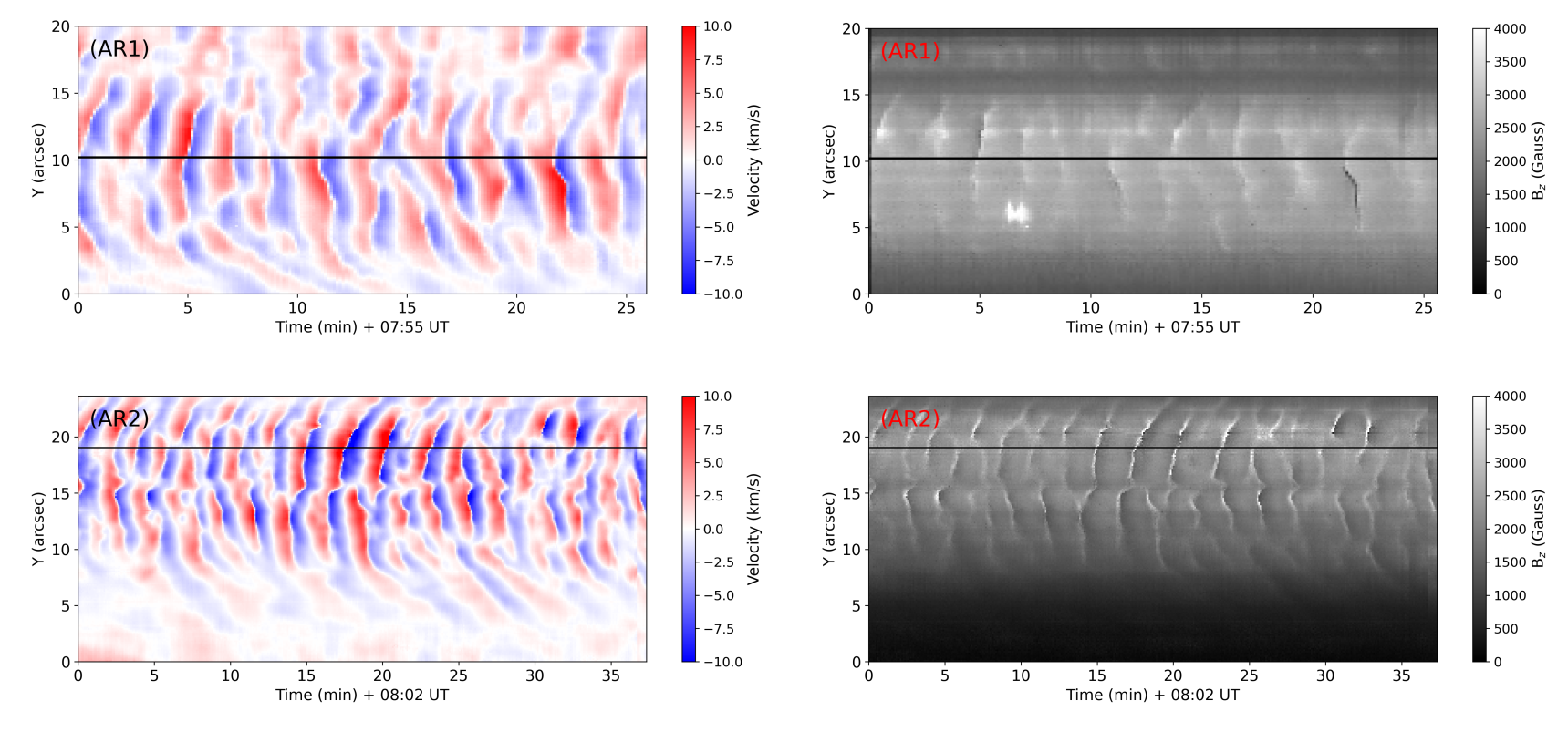}
	\\[0.1cm]	
	\includegraphics[width=0.99\linewidth]{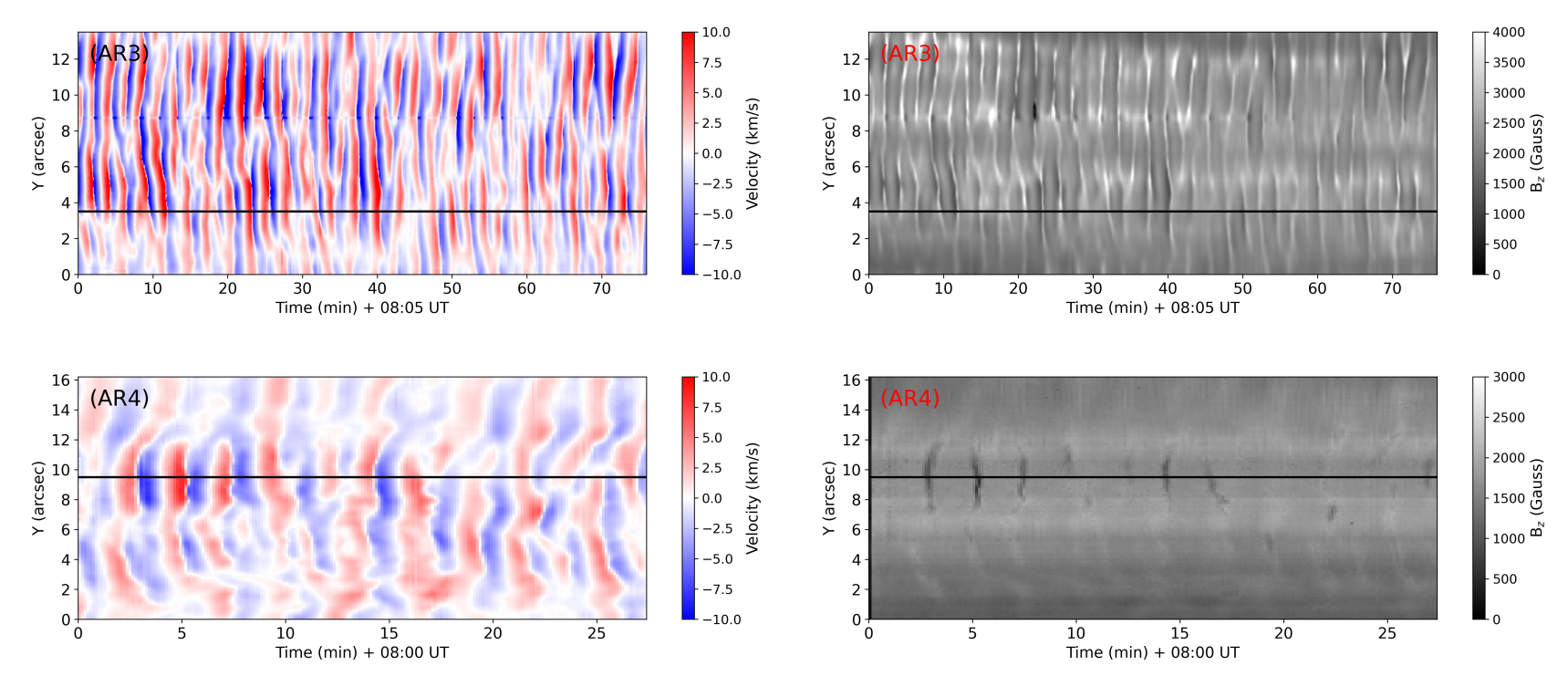}	
	\\[0.1cm]
	\includegraphics[width=0.99\linewidth]{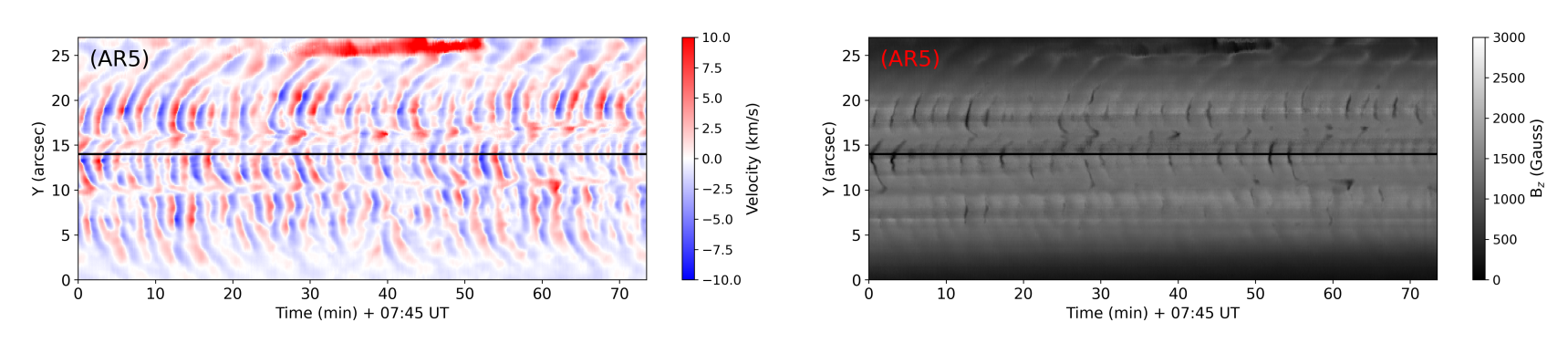}
	
	\caption{Temporal evolution of the chromospheric line-of-sight velocities (left panels) and line-of-sight magnetic fields (right panels) along the spectrograph slit for AR1, AR2, AR3, AR4 and AR5, respectively, inferred from the inversion of chromospheric He I triplet at 10830 {\AA} line. Horizontal black solid lines indicate the locations illustrated in Figure \ref{AR12345_Vlos_Bz_Temp}.}

	\label{AR12345_Bz_Vlos_Map}%
\end{figure*}

\begin{figure}
	\centering
	\includegraphics[width=0.85\linewidth]{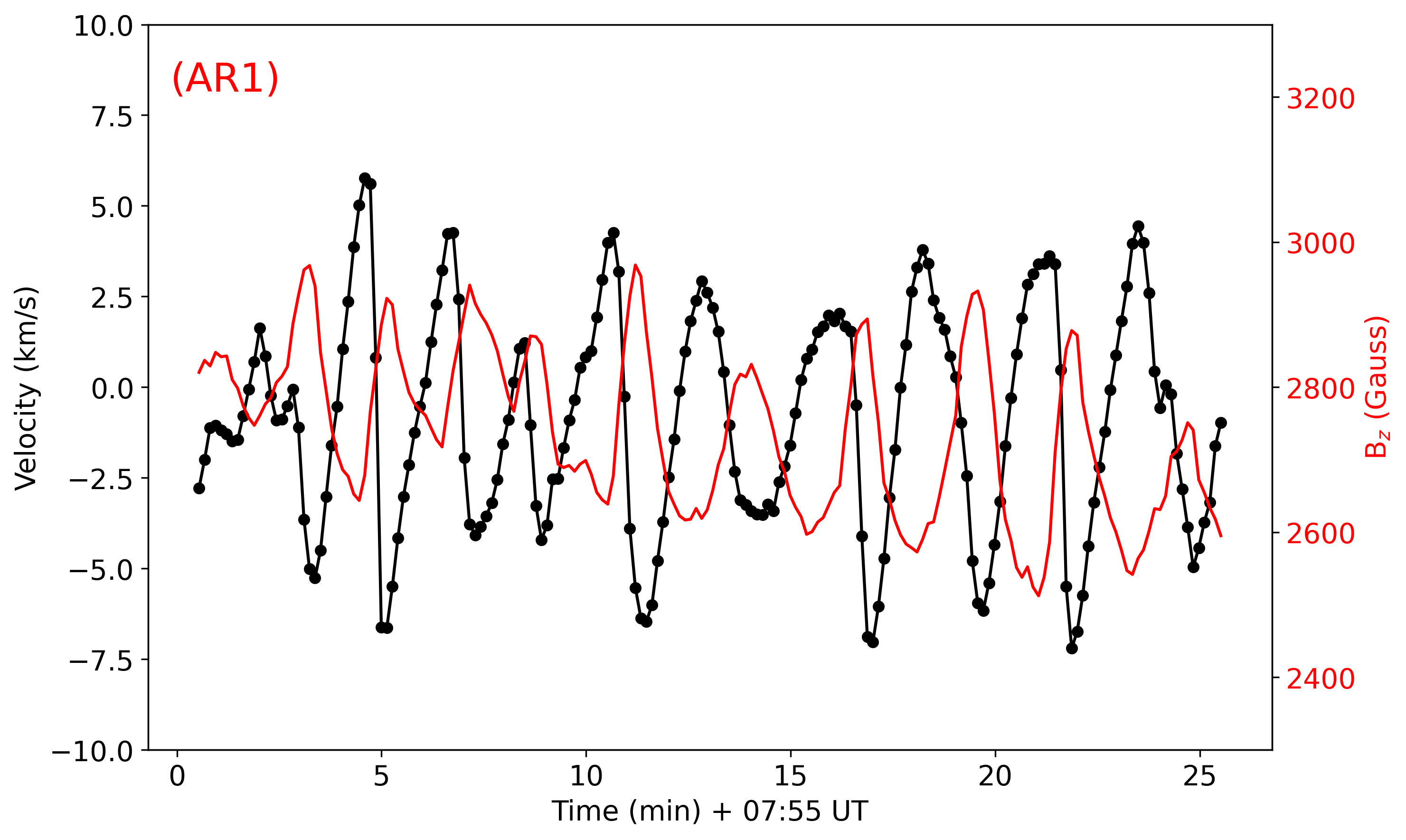}
	\\[0.1cm]	
	\includegraphics[width=0.85\linewidth]{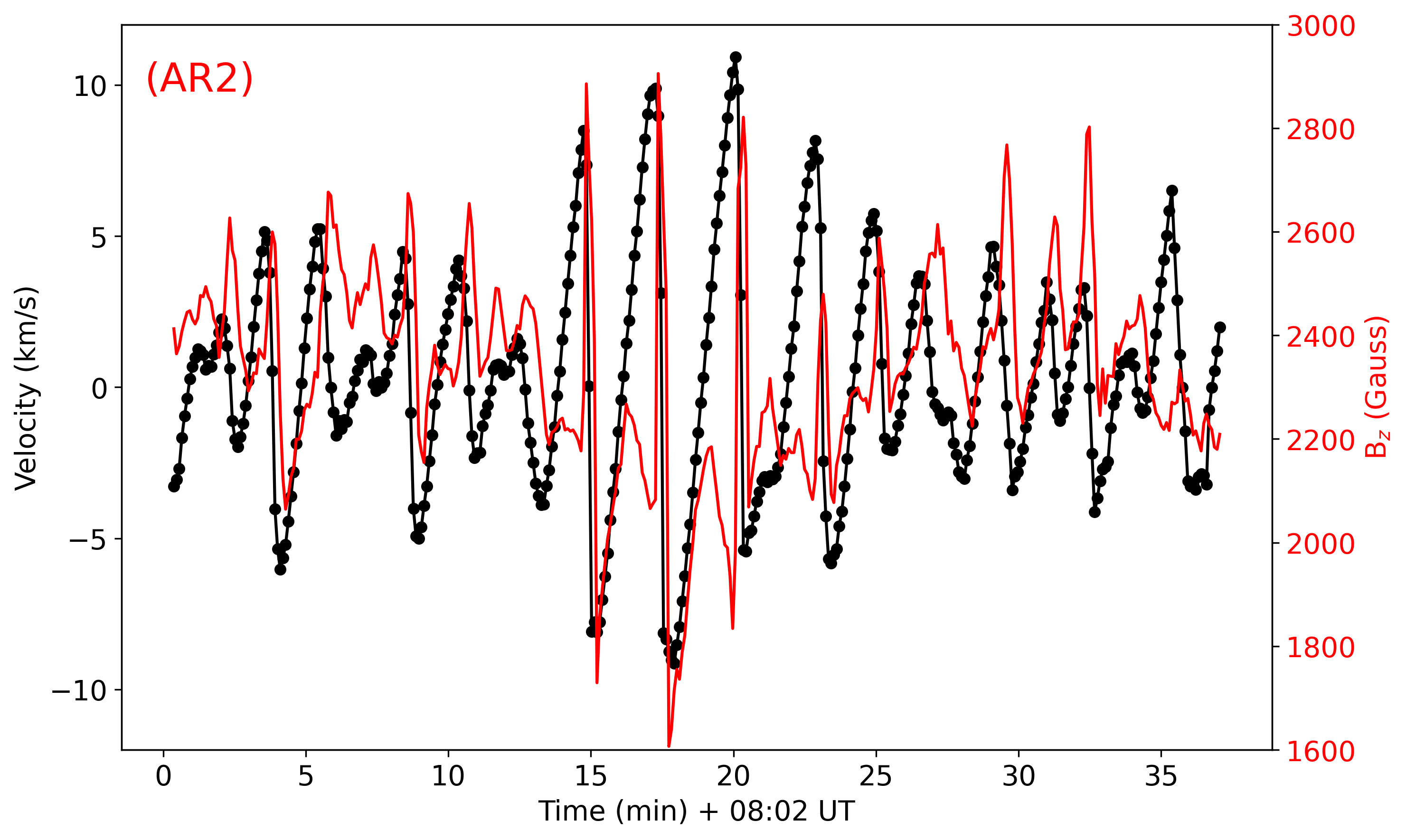}	
	\\[0.1cm]
	\includegraphics[width=0.85\linewidth]{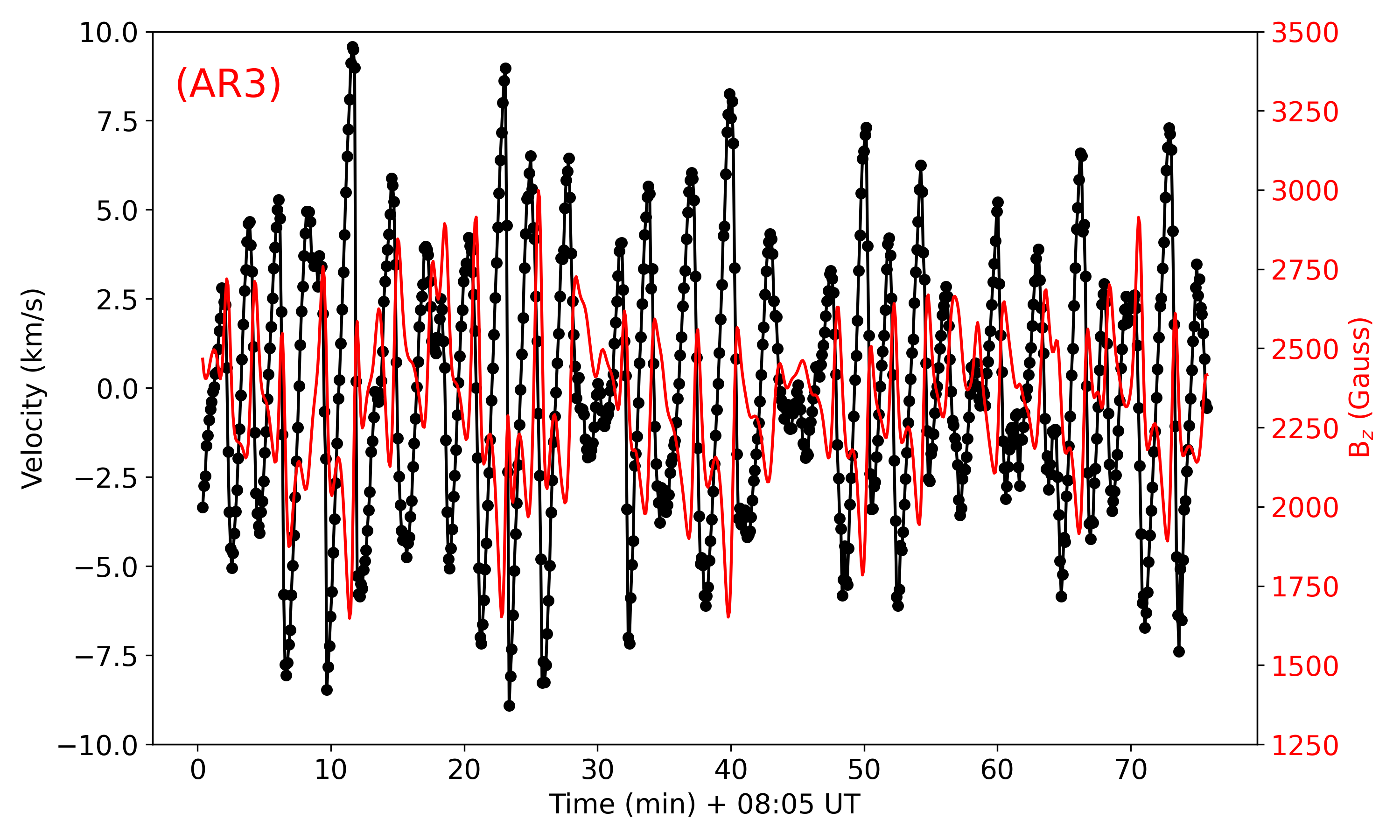}	
	\\[0.1cm]
	\includegraphics[width=0.85\linewidth]{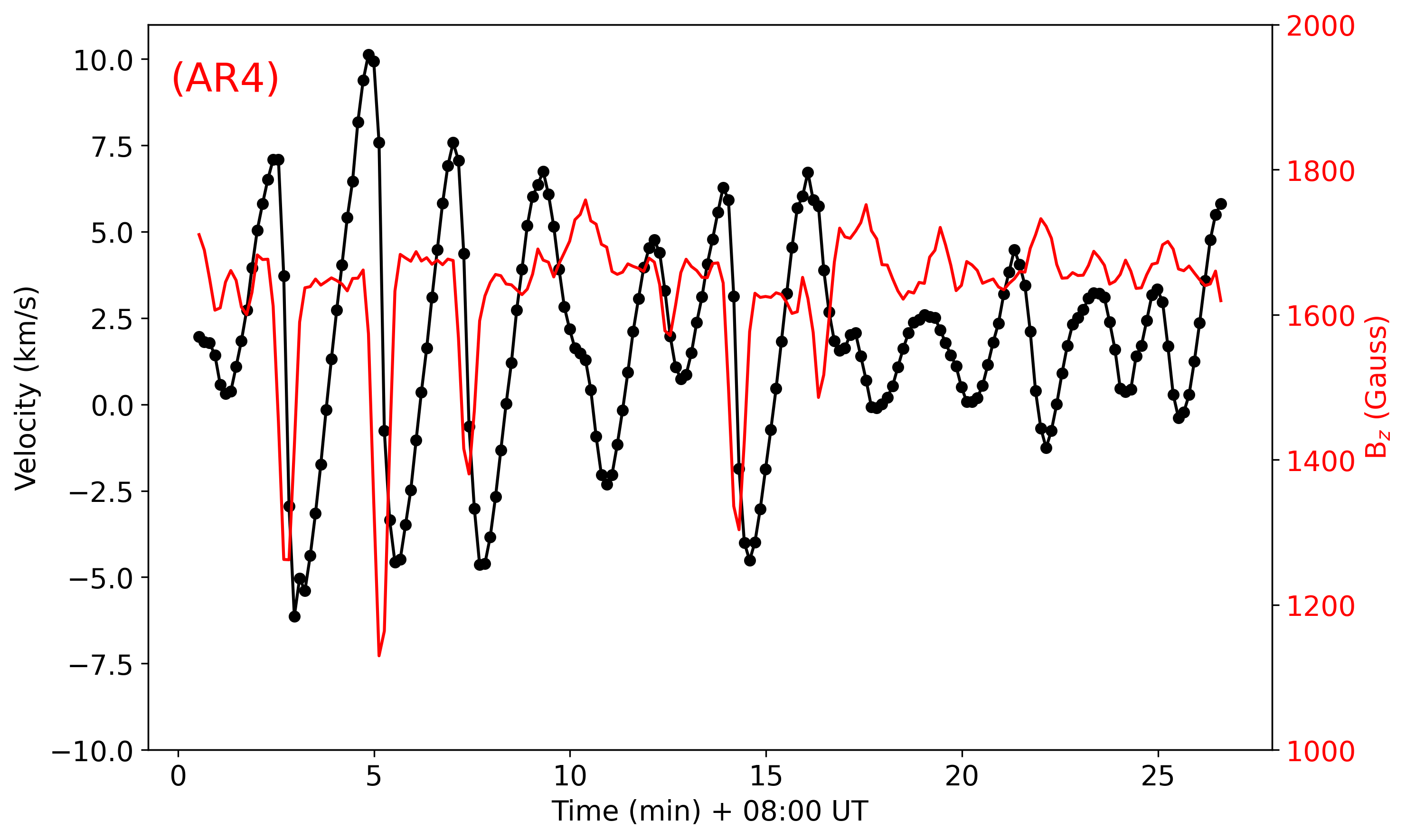}
	\\[0.1cm]
	\includegraphics[width=0.85\linewidth]{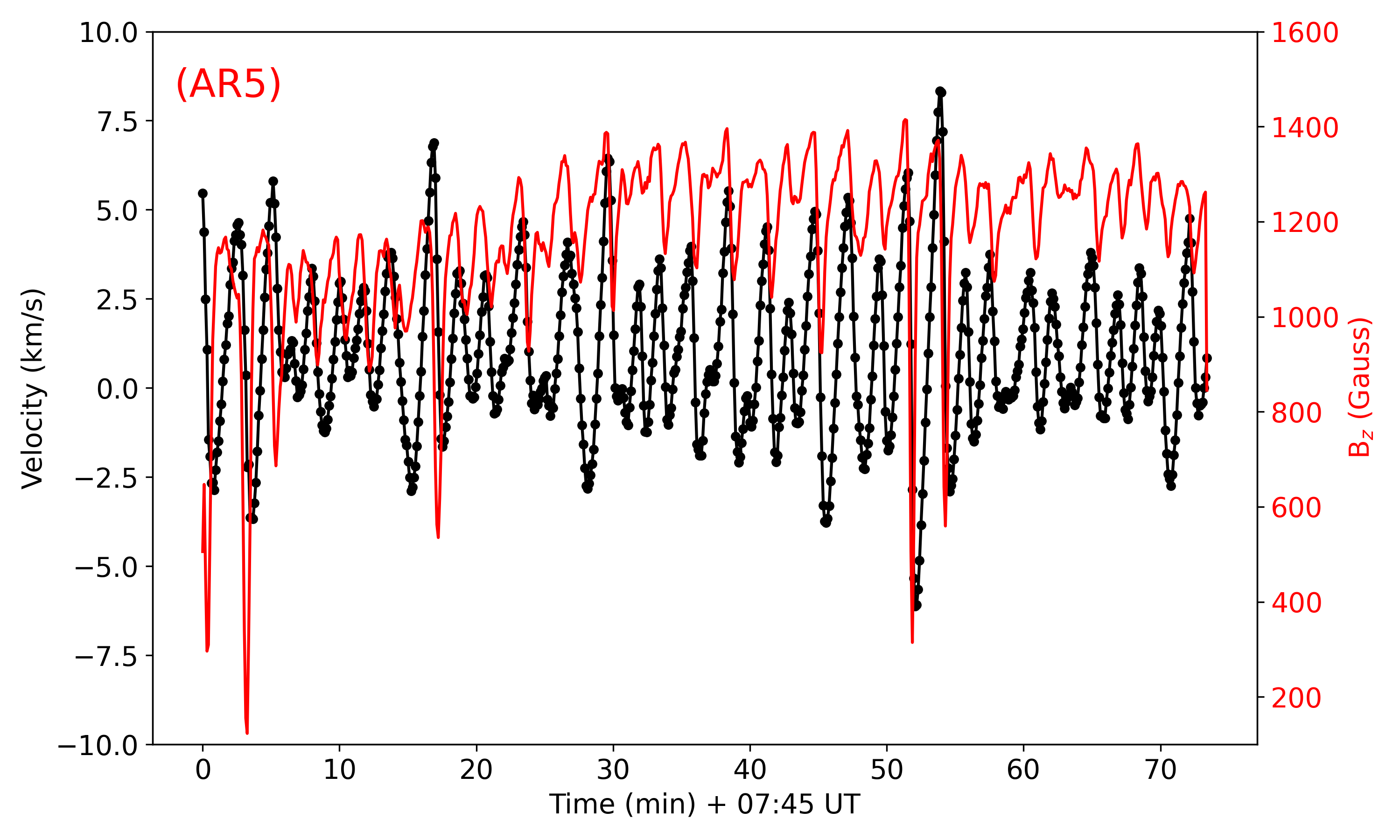}	
	
	\caption{Temporal evolution of the chromospheric line-of-sight velocity (black) and line-of-sight magnetic field (red) at selected locations indicated by black solid lines in Figure \ref{AR12345_Bz_Vlos_Map} and diamonds in Figure \ref{AR12345_HiFi}. In case of AR1, AR2 and AR3, magnetic field fluctuations have been smoothed by averaging in three time step windows. Each panel correspond to a different active region, from AR1 (top panel) to AR5 (bottom panel).}
	\label{AR12345_Vlos_Bz_Temp}
\end{figure}

Figure \ref{AR12345_Bz_Vlos_Map} illustrates the chromospheric line-of-sight velocity and magnetic field maps of all five sunspots inferred from 1-component inversions for the umbra and surroundings of the sunspots (white line in Figure \ref{AR12345_HiFi}). The velocity maps show the usual shock signatures as sudden changes of velocity, i.e., down (red) to up (blue) flows. This is also clearly seen as a saw tooth pattern in the temporal evolution of the velocity at selected pixels (Figure \ref{AR12345_Vlos_Bz_Temp}). We note a steady increase in velocity, followed by a sudden change from down to upflows. In all five sunspots, the amplitude of chromospheric velocity fluctuations is around 10 km s$^{-1}$. This is comparable to the local sound speed in the solar chromosphere, which is an essential element for shock development. The temporal evolution of the line-of-sight magnetic fields at the same pixels (red lines in Figure \ref{AR12345_Vlos_Bz_Temp}) exhibits striking behaviors. We found strong coherence between velocity oscillations and magnetic field fluctuations in the results for AR1, AR2, and AR3 (see the top three panels of Figure \ref{AR12345_Vlos_Bz_Temp}), with the magnetic field lagging velocity oscillations. In these three cases, the magnetic field increases up to 3000 G during or immediately after the shock. However, in the case of AR4 and AR5 (bottom two panels of Figure \ref{AR12345_Vlos_Bz_Temp}), a different magnetic field evolution pattern is observed during different phases of the magnetoacoustic shocks. We found an abrupt decrease in magnetic fields during the strongest shocks, whereas fluctuations of around 100 G occurs otherwise. Our study focuses on magnetic field fluctuations accompanying magnetoacoustic shocks. Therefore, the analysis is restricted to umbral regions.

\begin{figure}
	\centering
	
	\includegraphics[width=0.89\linewidth]{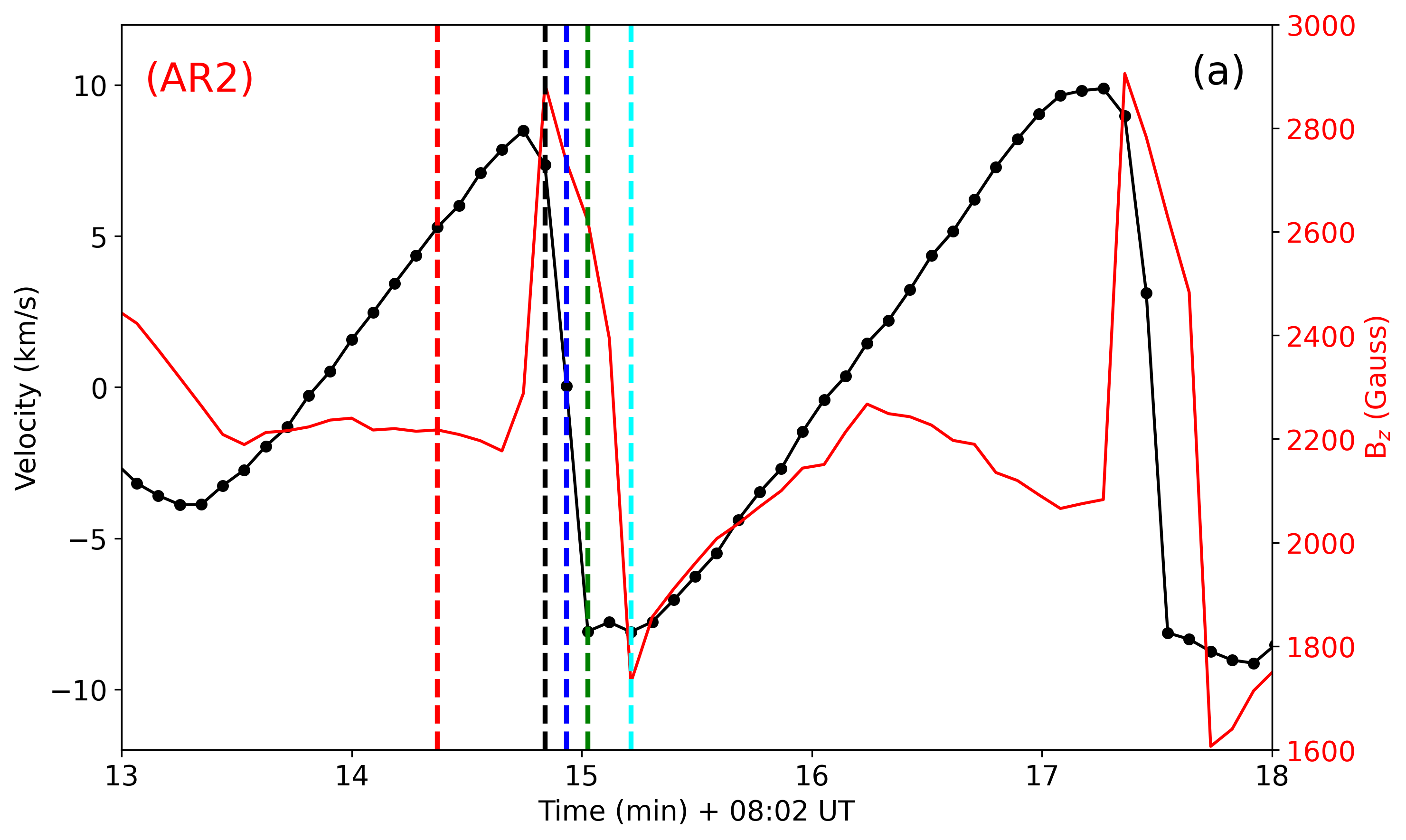}
	\includegraphics[width=0.89\linewidth]{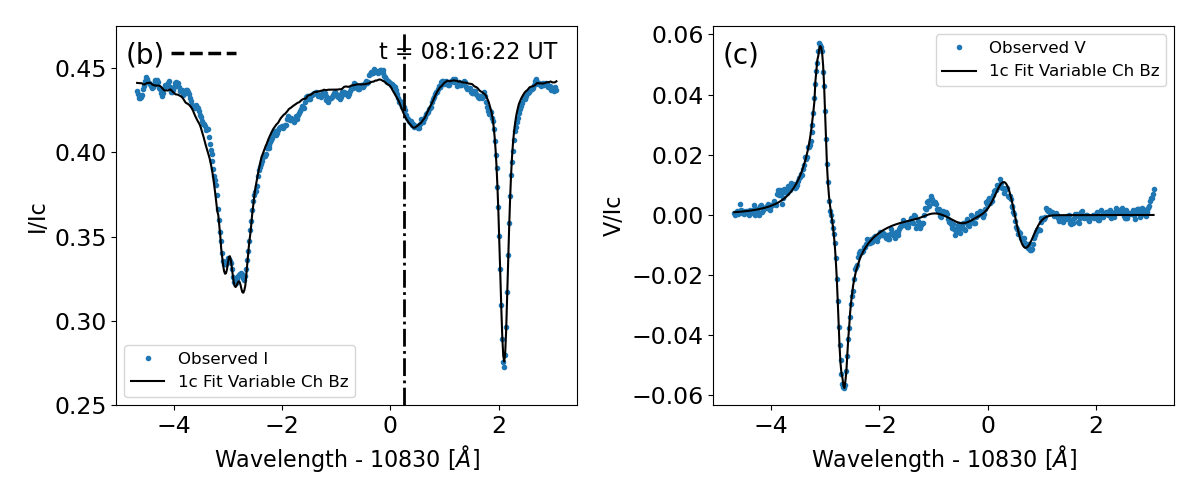}
	\includegraphics[width=0.89\linewidth]{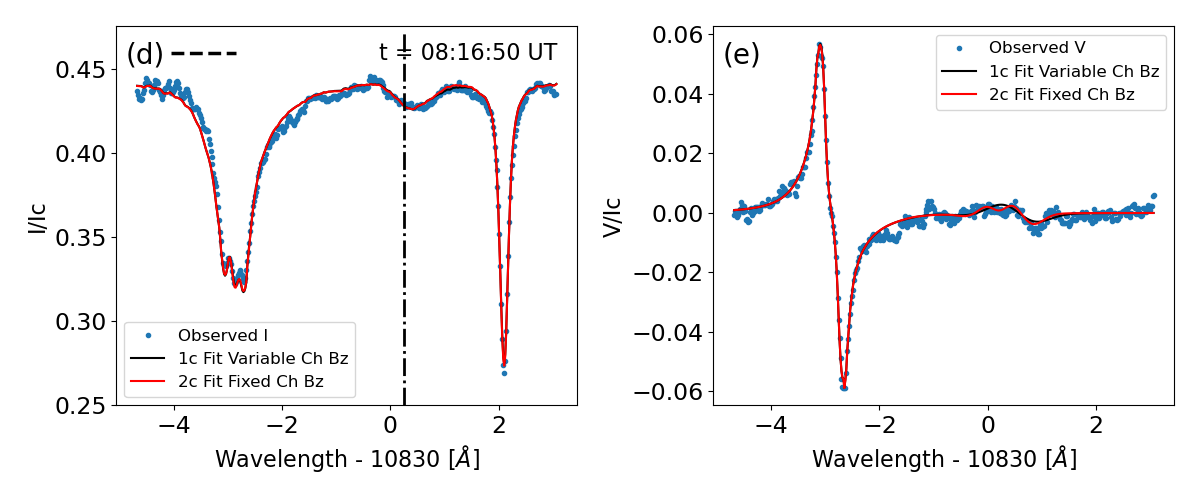}
	\includegraphics[width=0.89\linewidth]{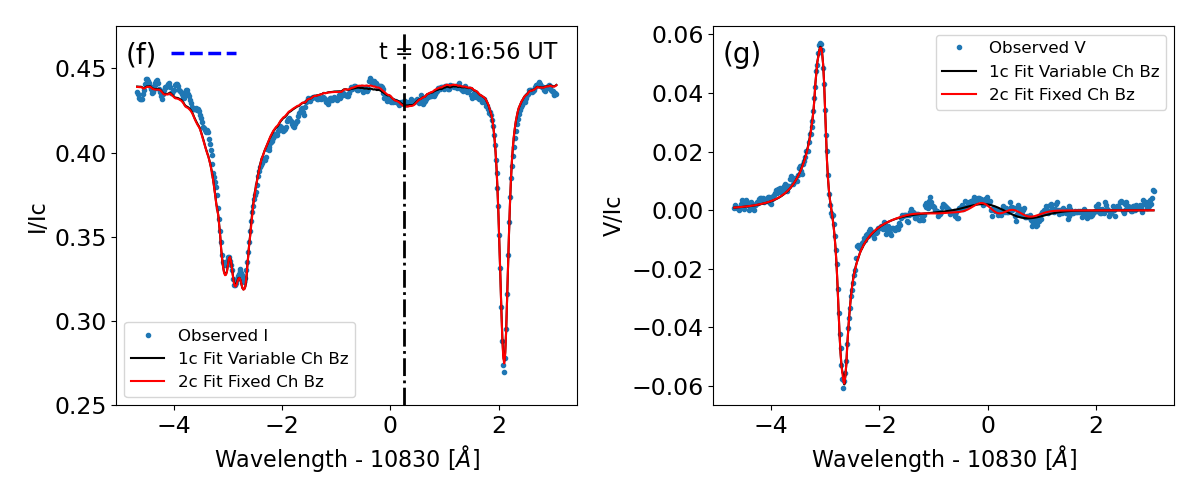}
	\includegraphics[width=0.89\linewidth]{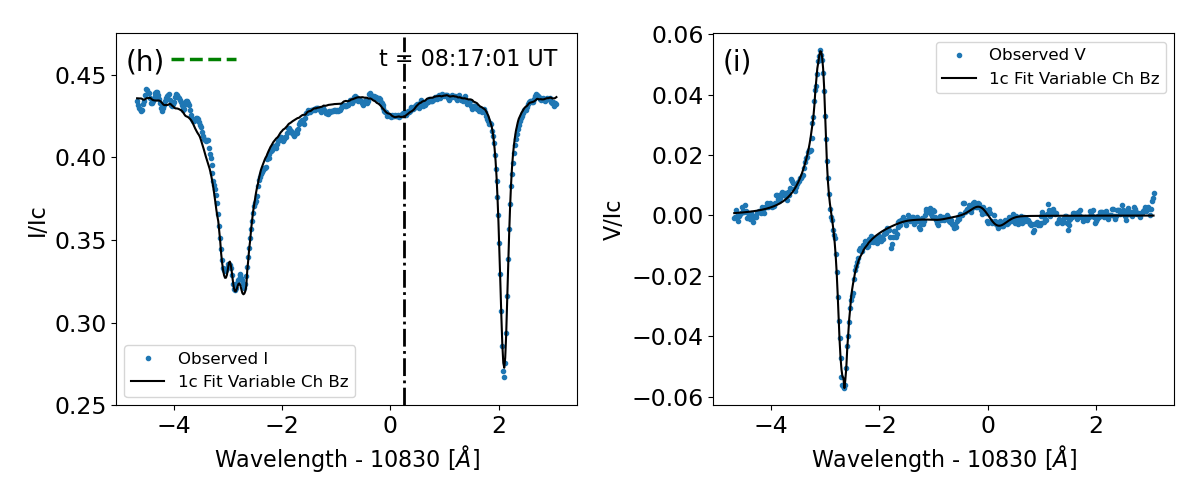}	
	\includegraphics[width=0.89\linewidth]{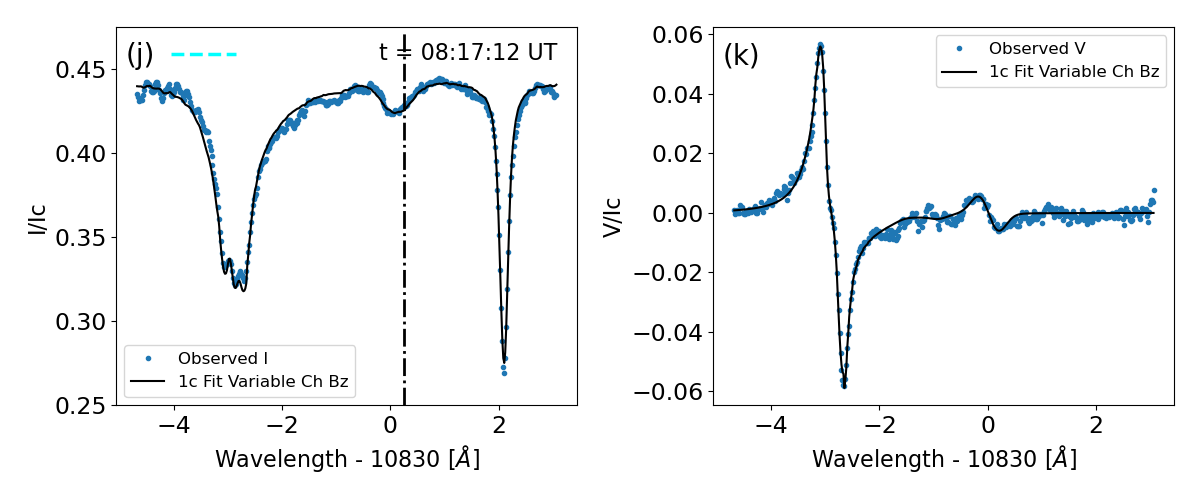}
	
	\caption{Top panel: Zoom of the temporal evolution of the chromospheric line-of-sight velocity (black) and line-of-sight magnetic field (red) from AR2 (second panel from Figure \ref{AR12345_Vlos_Bz_Temp}). Lower panels show the observed Stokes I and V spectral profiles (blue dots) and the fits from 1-component variable magnetic field (black lines) and 2-component with fixed chromospheric magnetic fields (red lines) inversions, at the times indicated by the red, black, blue, green and cyan dashed vertical lines in the top panel (a) as well as in the left panels. The black dot-dashed vertical line in panels (b), (d), (f ), (h), and (j) highlight the rest wavelength at 10830.25 {\AA}.}	
	\label{AR2_Spectral_Profiles}
	
\end{figure}

\begin{figure}
	\centering
	
	\includegraphics[width=0.92\linewidth]{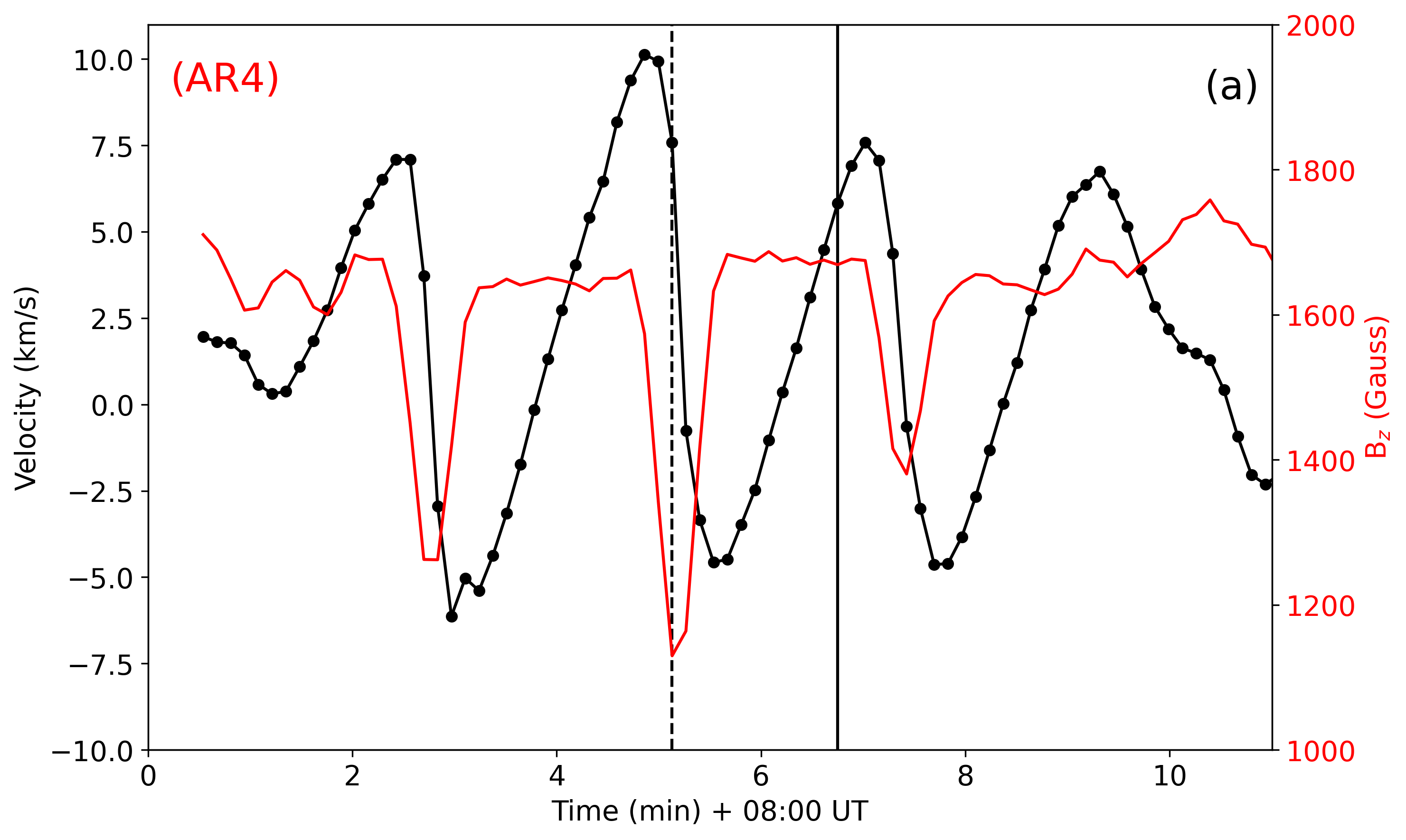}	
	\includegraphics[width=0.92\linewidth]{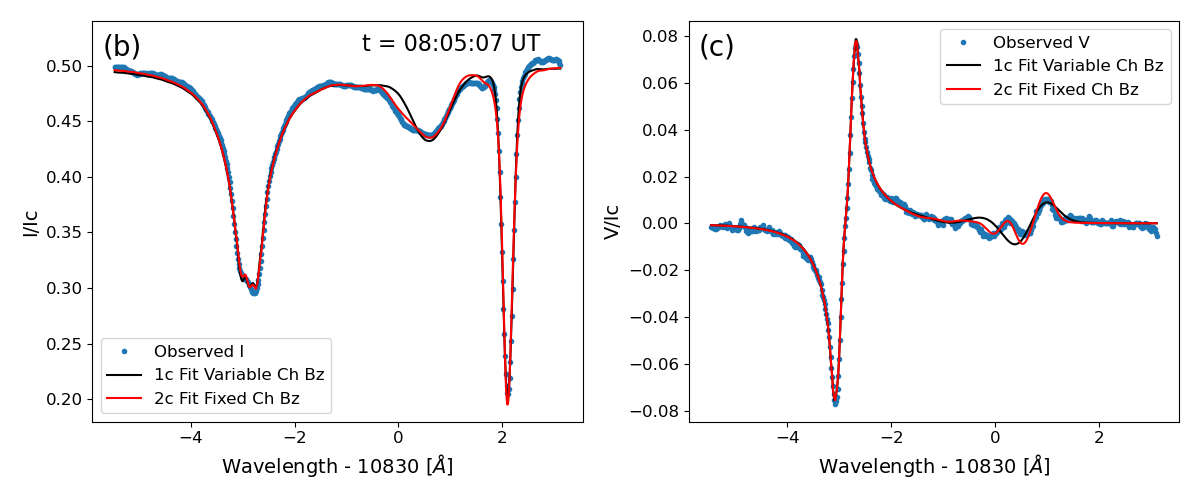}
	\includegraphics[width=0.92\linewidth]{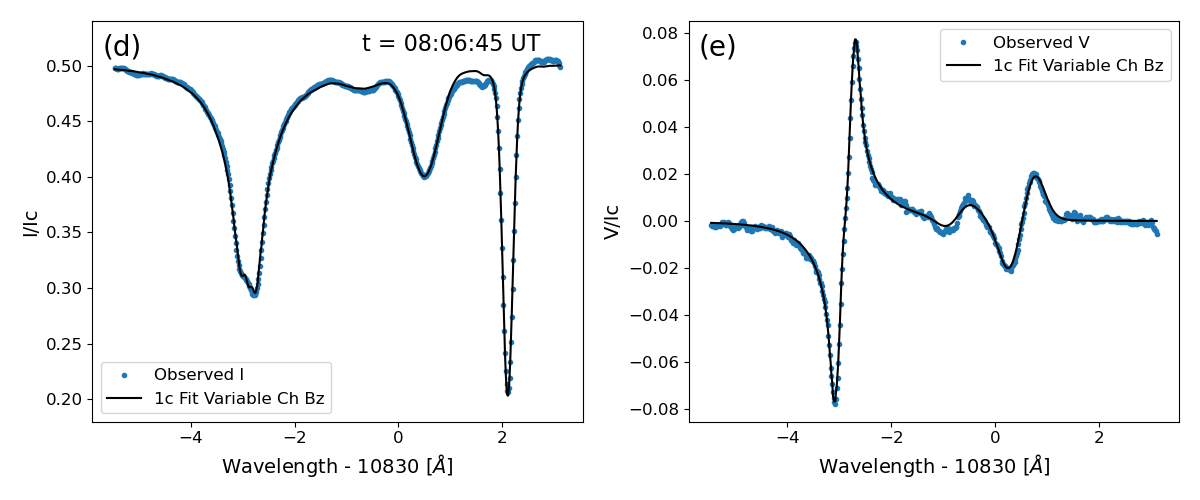}
	\caption{Same as Figure \ref{AR2_Spectral_Profiles} but for AR4. Middle panels illustrate the shock (vertical dashed line in panel a) and bottom panels the quiescent state (vertical solid line in panel a).}
	
	\label{AR4_Spectral_Profiles}
	
\end{figure}

\begin{figure}
	\centering
	\includegraphics[width=0.92\linewidth]{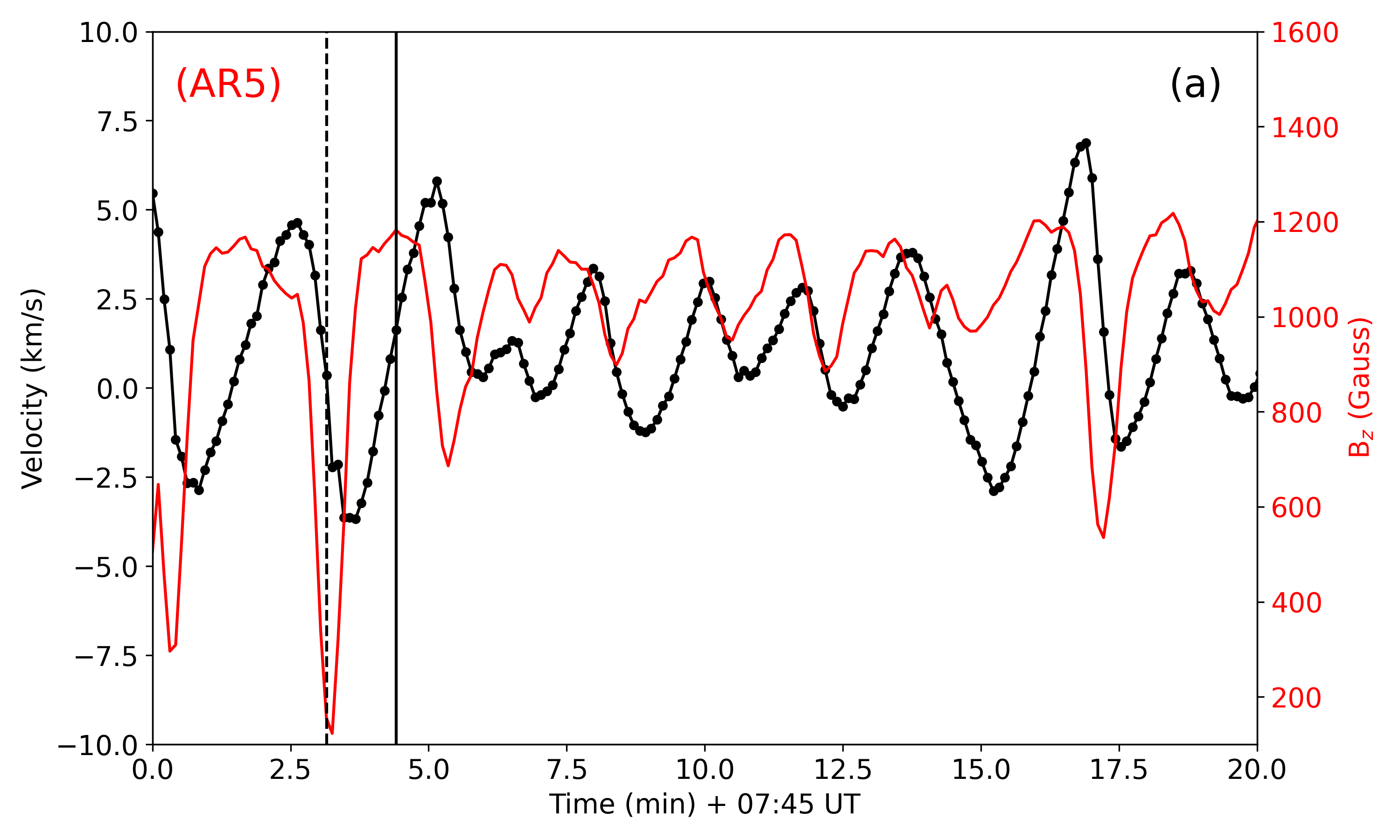}
	\includegraphics[width=0.92\linewidth]{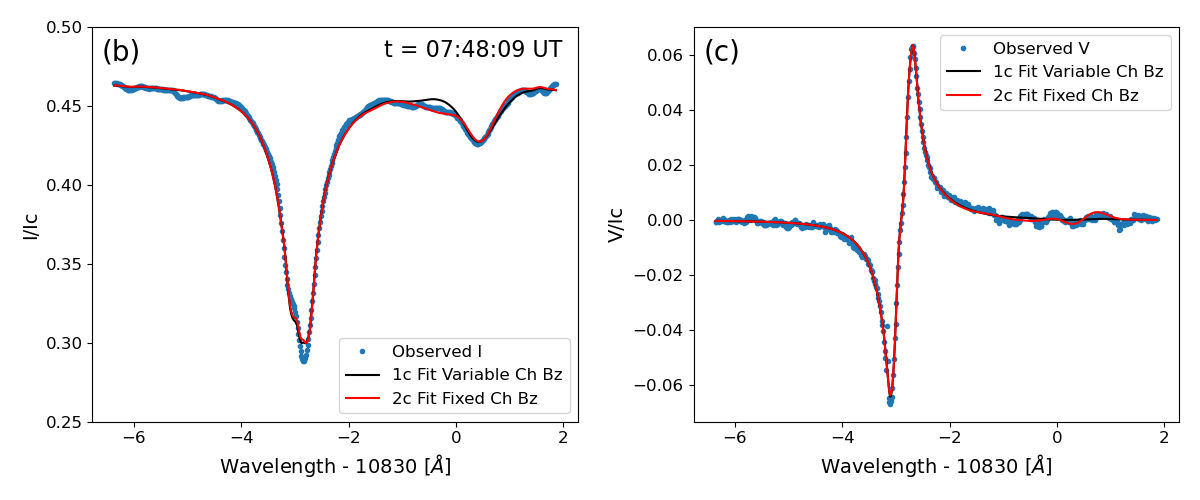}
	\includegraphics[width=0.92\linewidth]{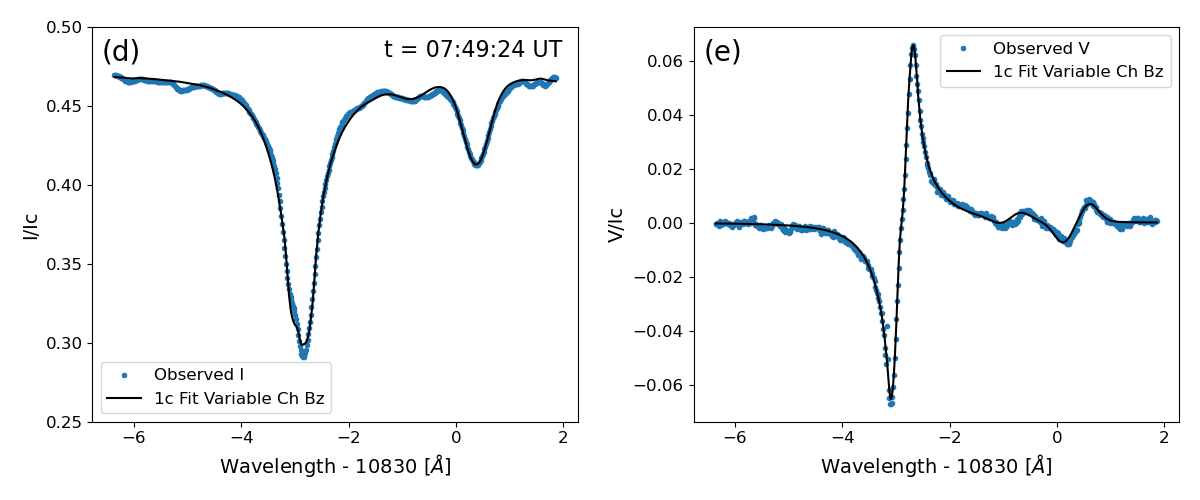}
	
	\caption{Same as Figure \ref{AR2_Spectral_Profiles} but for AR5. Middle panels illustrate the shock (vertical dashed line in panel a) and bottom panels the quiescent state (vertical solid line in panel a).}
	
	\label{AR5_Spectral_Profiles}
\end{figure}

\begin{figure}
	\centering
	\includegraphics[width=0.92\linewidth]{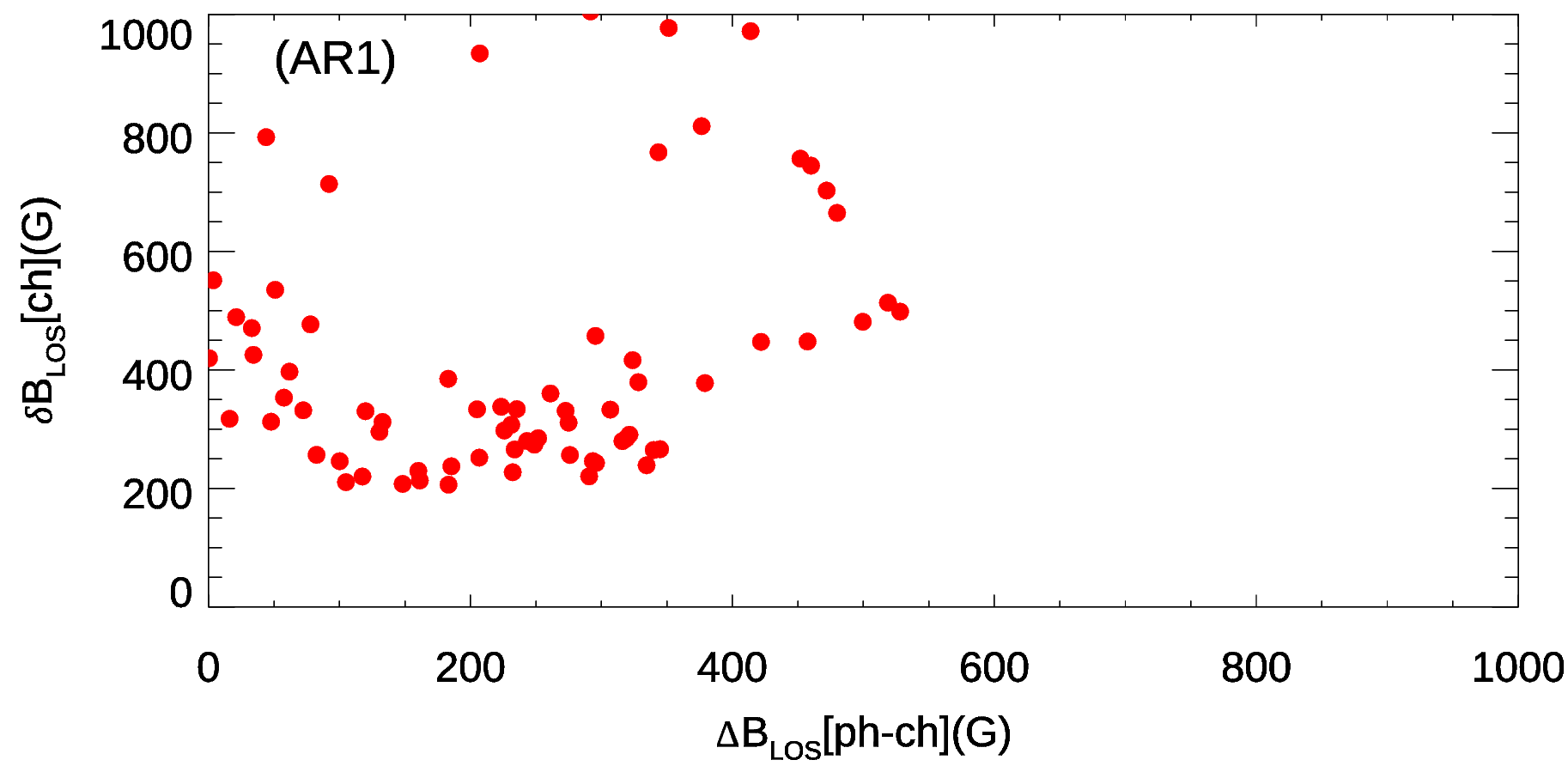}
	\includegraphics[width=0.92\linewidth]{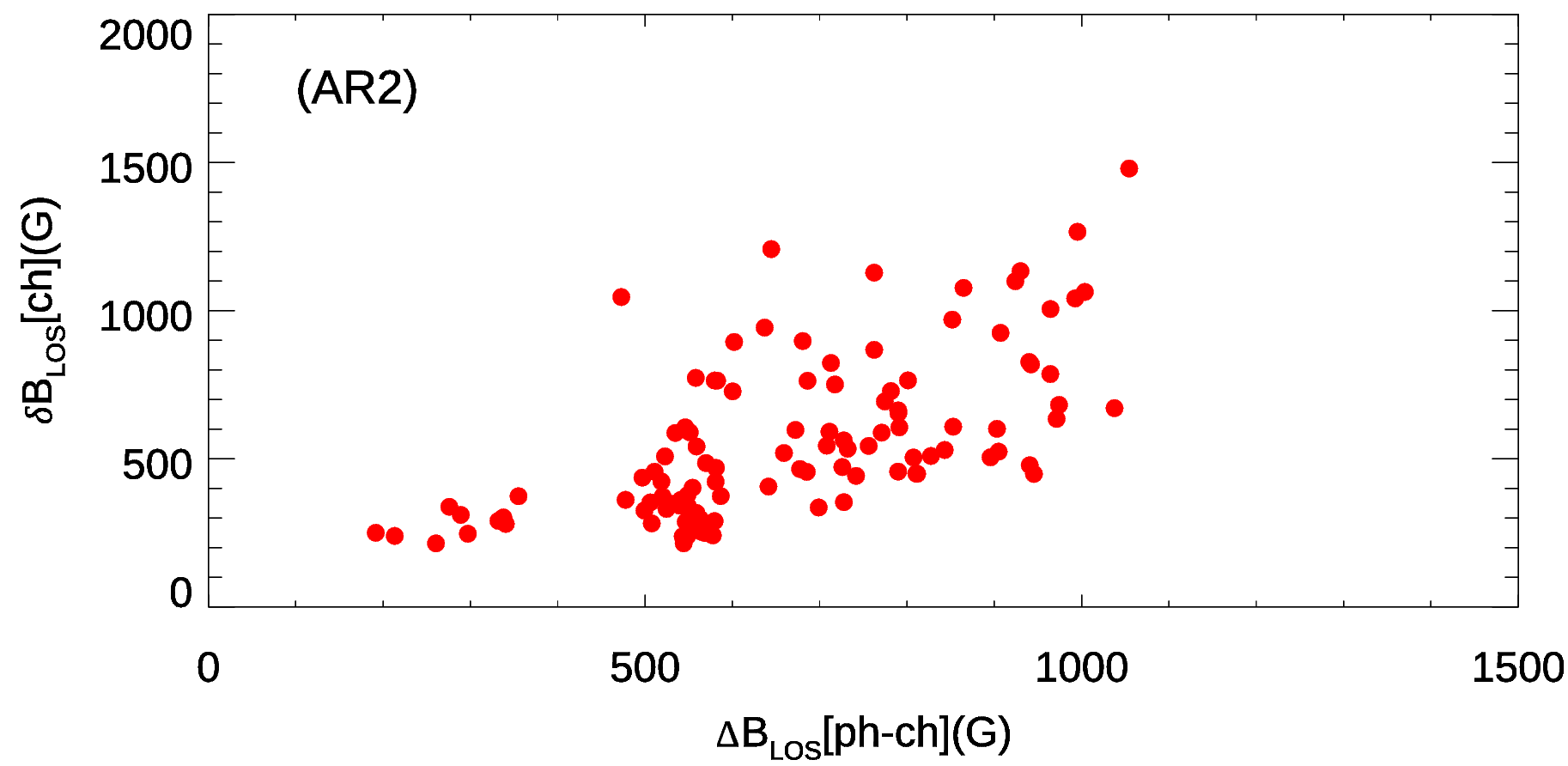}	
	\includegraphics[width=0.92\linewidth]{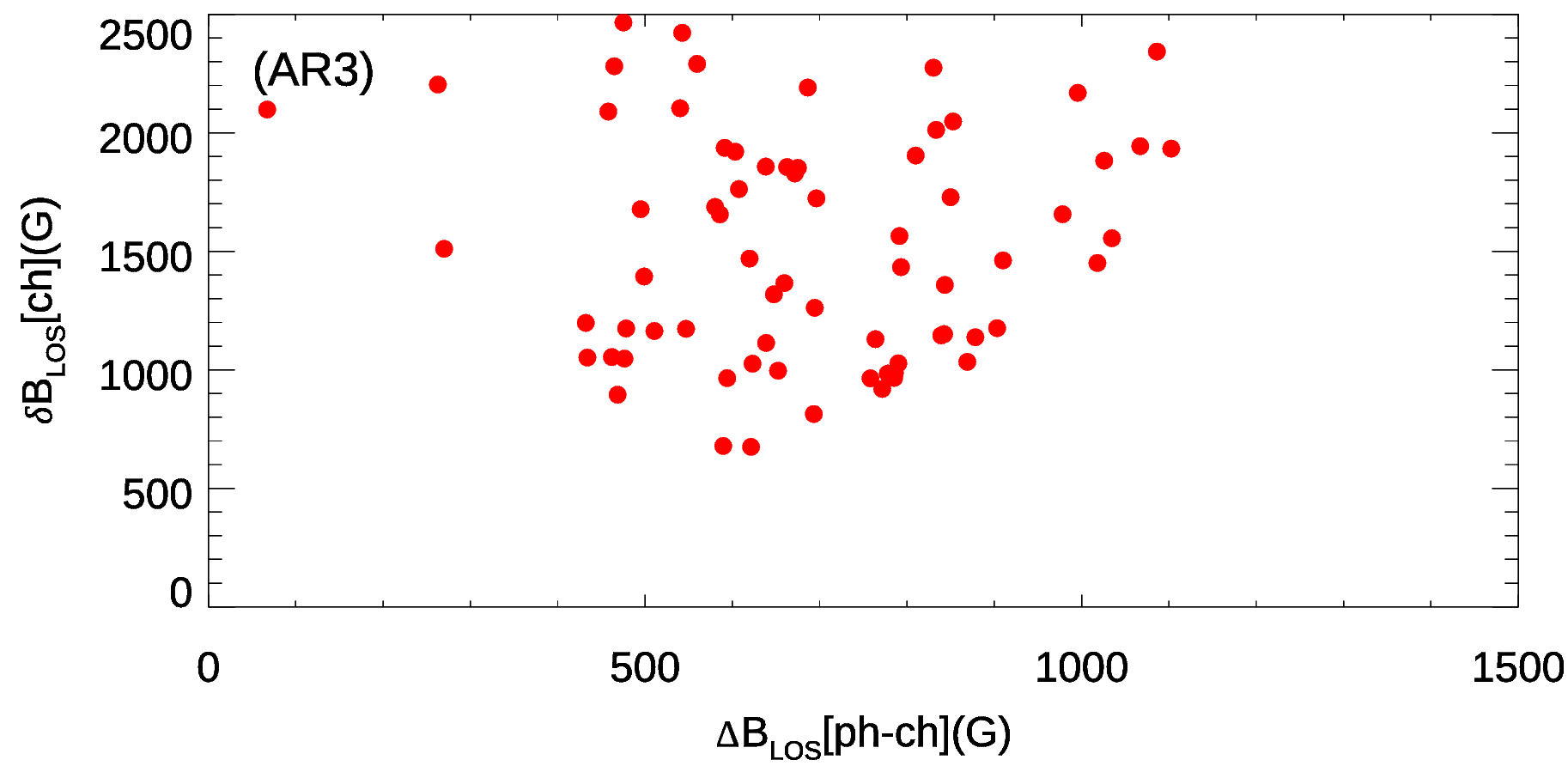}
	\caption{Scatter plots of the gradients between photosphere and chromosphere i.e. $\Delta B_{LOS}[ph-ch]$ (G) and chromospheric magnetic fields fluctuations i.e. $\delta B_{LOS}[ch]$ (G) for the AR1, AR2 and AR3, respectively.}
	\label{AR123_Bz_gradients}
\end{figure}

\begin{table*}
	\caption{Details of the outcome from the inversions using 1- and 2 component chromosphere for AR5 as shown in the panels (b) and (c) from Figure \ref{AR5_Spectral_Profiles}. In the 2-component inversion, the chomospheric magnetic field (B$_{z}$) was fixed to -1100 G.}
	\label{2comp}
	\centering
	
\begin{tabular}{l c c c c c c c |c |c }
	\hline
	& Bz (G) & $\tau$ & V$_{LOS}$ (km s$^{-1}$) & $\Delta$v (km s$^{-1}$) & $\beta$ & a & ff & BIC(I) & BIC(V) \\
	\hline
	1 c & -175.399 & 0.086 & 4.0 & 14.0 & 1.0 & 0.1 & 1.0 & 8903.04 & 2237.62 \\
	\hline
	2c, ch1 & -1100.0 & 0.057 & -14.006 & 12.0 & 1.0 & 0.0 & 1.0 & \multirow{2}{*}{3941.69} & \multirow{2}{*}{2012.01} \\
	\cline{1-8}
	2c, ch2 & -1100.0 & 0.145 & 5.257 & 8.905 & 1.675 & 0.1 & 1.0 & & \\
	\hline
\end{tabular}

\end{table*}

\subsection{Examination of spectral profiles during shocks}
\label{sect:profiles}

Figure \ref{AR2_Spectral_Profiles} shows the Stokes I and V profiles accompanying one of the shocks from AR2, displaying temporal variations that are closely associated with different phases of the magnetoacoustic shock. The temporal evolution of the chromospheric velocity and magnetic field, derived from the 1-component inversion around the time of the spectrum, is also included for context. 

The first example profiles (panels b and c) illustrate the spectra during the progressive phase of the velocity evolution, as highlighted by the red dashed vertical line (Figure \ref{AR2_Spectral_Profiles} (a)), prior to the occurrence of a shock. The 1-component inversion provides a good fit of the observed Stokes profiles. Discrepancies are found in the core of the Si I 10827 {\AA} line since it is sensitive to NLTE effects \citep{2012KPCB...28..169S, 2017A&A...603A..98S, 2019A&A...628A..47S}. In our analysis, we have treated the Si I line in LTE. The photospheric magnetic field employed in the following analyses is based on the inversion results at log{$\tau$} = -1.5. \cite{2012A&A...539A.131K} compared NLTE and LTE effects of the same Si I line using departure coefficients within SIR. They concluded that NLTE effects mainly affect the inferred temperature, but only show small variations for the magnetic field and line-of-sight velocity for log tau = -2. Recently, the comparative analysis performed by \cite{2024A&A...692A.169Q} of the LTE and NLTE approximation for the Si I line highlights that, up to log{$\tau$} = -2, LTE provides an accurate characterization of the spectral line. 

During the development of the shock, chromospheric plasma undergoes a rapid change from downflow to upflow and localized heating. The signature of these changes is imprinted in the Stokes profiles. A sudden shift in the core of the He I 10830 {\AA} line from redshifted (panel d) to blueshifted (panels f and h) is clearly visible. Also, the depth of the line is reduced during this stage of the shock, which is accompanied by a significant reduction in the amplitude of the Stokes V signal. The 1-component inversion can reproduce these profiles by accounting for the changes in the plasma velocity, from 8 km s$^{-1}$ at the beginning of the shock (black dashed line in panel a, profiles in panels d and e) to -8 km s$^{-1}$ at the end of the shock (green dashed line in panel a, profiles in panels h and i). Interestingly, the line-of-sight magnetic field inferred from these inversions exhibits a remarkable enhancement in the field strength, with a perturbation up to 600 G. Alternatively, we have also fit the profiles at the shock stage using a model with two chromospheric components and a fixed magnetic field at 2300 G. This model improves the fit of Stokes V in the central part of the shock (blue dashed line in panel a, Stokes V in panel g), providing a better match to the multi-lobe pattern found in circular polarization in He I 10830 {\AA}. In contrast, at the next time step (panels h and i) the 1-component inversion gives a better fit of the intensity in He I 10830 {\AA} and a similarly good match in Stokes V.

After the shock (cyan dashed line in panel a, Stokes profiles in panels j and k), the line depth and Stokes V amplitude start to increase again. In this case, the 1-component inversion fits the profiles with a significant reduction in the magnetic field (around 500 G below the quiescent value). The 2-component inversion can also fit these profiles without introducing magnetic field fluctuations. All in all, we find that in this sunspot, both inversion strategies (accounting for magnetic field variation in the case of the 1-component inversion or strong velocity gradients in the case of the 2-component inversions) can reproduce the observed profiles. This conclusion applies to the active region illustrated in Figure \ref{AR2_Spectral_Profiles} (AR2), but also to AR1 and AR3 (spectral profiles not included in the paper, see magnetic field fluctuations from the 1-component inversion in Figure \ref{AR12345_Vlos_Bz_Temp}).

Figures \ref{AR4_Spectral_Profiles} and \ref{AR5_Spectral_Profiles} also illustrate some selected profiles from two of the observed sunspots, AR4 and AR5, respectively. Both figures show example profiles at two different time steps, marked with vertical lines in the top panel, one of them prior to the shock and the other during the shock. As previously pointed out, the temporal evolution of the magnetic field inferred from the 1-component inversions of these two active regions (AR4 and AR5) shows a much different behavior. Instead of magnetic field enhancements, the shocks are associated with dramatic reductions in the field strength, reaching values even below 200 G in the case of AR5. 

An examination of the spectral profiles reveals that, during the shocks, an inversion strategy with 1-component is not able to properly model the observations. The intensity profiles of the He I 10830 {\AA} at the shock stages (panels b from Figures \ref{AR4_Spectral_Profiles} and \ref{AR5_Spectral_Profiles}) exhibits clear insights of a multi-component atmosphere, as indicated by the asymmetric line widths. Also, the 1-component inversion cannot capture the several lobes found in Stokes V. In contrast, 2-component inversions provide a much better reproduction of the observed profiles at those stages of the chromospheric oscillation. Our analysis also shows that, out of the shock stages, the 1-component inversions gives an accurate characterization of the observed Stokes profiles. Table \ref{2comp} illustrate the results obtained from the 1- and 2-component inversion of the profiles during the shock from AR5 as shown in the panels b and c from Figure \ref{AR5_Spectral_Profiles}. Using 1-component inversion, the inferred magnetic field was around 175 G, whereas 2-component inversions can better reproduce the same Stokes V profiles with a 1100 G fixed chromospheric magnetic fields through changes in the line-of-sight velocity and width of the profile (see  Figure \ref{AR5_Spectral_Profiles} c). In the 2 component inversion, a very strong up flows of the order of around 14 km s$^{-1}$ has been found in one of the chromospheric components, whereas the other component exhibits a 5 km s$^{-1}$ downflow. Although the improvement in the fitting of the Stokes I and V signals achieved when using 2-components is visually noticeable, we have quantified it using the Bayesian Information Criterion (BIC: \citealt{1978AnSta...6..461S, 2012ApJ...748...83A}). It is estimated with the following expression: BIC = $\chi^{2}$ + k*log(N), where, k is number of free parameter in a model and N is number of frequency points used for the fitting. We have computed it over a wavelength range of $\pm$1.5 {\AA} from the center of He I 10830 {\AA}. We found that BIC is smaller in the fitting of the 2-component models compared to the 1-component, favoring the 2-component inversion strategy for these profiles.

\section{Discussion}
\label{Sect4}

We have analyzed the He I 10830 {\AA} spectropolarimetric signals during magnetoacoustic shocks in a sample of five sunspots. In three cases, the Stokes profiles can be interpreted as the result of remarkable fluctuations in the line-of-sight magnetic field or the contribution from two chromospheric slabs with different velocities. In the other two sunspots, the 1-component inversions retrieve solutions with striking reductions in the magnetic field. However, the spectra from these solutions provide a poor fitting of the observed Stokes profiles and, thus, has been discarded. For these two sunspots, only models with two chromospheric components can actually explain the observations.  \\

Here, we discuss two scenarios that can explain the spectropolarimetric signals during the passage of the shocks.

\subsection{Magnetic fields and fluctuations in the response height of He I 10830 {\AA}} 

The estimated higher magnetic fields in the umbral chromosphere, reaching up to 3000 G in AR1, AR2, and AR3, are difficult to associate with the background chromospheric magnetic field. Previous independent analysis performed by \cite{2023A&A...676A..77F} of the He I 10830 {\AA} line reported chromospheric umbral magnetic fields of the order of 2900 G during magnetoacoustic shocks. They interpreted these fluctuations as remarkable changes in the response height of the He I 10830 {\AA} to the magnetic field. In this scenario, the He I 10830 {\AA} line probes the upper photosphere during the shocks. Other studies have reported similar variations in the response of the line associated with other events, such as flares \citep{2018PASJ...70..101A, 2019A&A...621A..35L} and supersonic downflows \citep{2021ApJ...916....5S}.

\cite{2023A&A...676A..77F} claim is supported by the linear relation observed between the amplitude of the inferred magnetic field fluctuations in the umbra ($\delta B_{LOS}[ch]$) and the gradient in the magnetic field between the photosphere and chromosphere ($\Delta B_{LOS}[ph-ch]$). Here, we have adopted the same approach and we have evaluated the same quantities for the three sunspots that exhibit magnetic field enhancements during shocks. One of these sunspots (AR2) corresponds to the active region analyzed in \cite{2023A&A...676A..77F}. The results presented in this study correspond to a completely independent analysis of the same data set. The main difference in the inversion scheme is that in this new study the filling factor of the stray light is fixed at every spatial location, whereas in the inversion scheme of \cite{2023A&A...676A..77F} temporal variations of the filling factor were allowed.

The details of the calculation of $\delta B_{LOS}[ch]$ and $\Delta B_{LOS}[ph-ch]$ are explained in \cite{2023A&A...676A..77F}. Figure \ref{AR123_Bz_gradients} illustrates the scatter plots of these quantities for AR1, AR2 and AR3. We have not found a clear correlation between the outcomes from AR1 and AR3 (see the top and bottom panels of Figure \ref{AR123_Bz_gradients}). In contrast, in the case of AR2, a linear correlation (Pearson correlation coefficient of 0.64) between $\delta B_{LOS}[ch]$ and $\Delta B_{LOS}[ph-ch]$ is found (see middle panel of Figure \ref{AR123_Bz_gradients}). This is the same active region from \cite{2023A&A...676A..77F}. The new independent analysis developed in this work show a qualitative agreement with the outcome from the previous study. The magnetic field fluctuations inferred from the 1-component inversions of AR2 are consistent with a lower response height of the He I 10830 {\AA} triplet during the shock. However, this scenario does not account for the reduction of the magnetic field strength during shocks (as seen in AR4 and AR5, Figures \ref{AR4_Spectral_Profiles} and \ref{AR5_Spectral_Profiles}, respectively) or the enhanced magnetic fields inferred for AR1 and AR3.

\subsection{Velocity gradients} 

Our results show that models with two chromospheric components not only can reproduce the Stokes profiles observed during the transit of shock waves, but also in some cases (Figures \ref{AR4_Spectral_Profiles} and \ref{AR5_Spectral_Profiles}) they are mandatory to reproduce the spectral signatures, as mentioned in Section \ref{sect:profiles}. In these 2-component inversions, we have chosen to set a fixed magnetic field strength for both chromospheres, which was determined at each location as the median value (over all temporal steps) from the previously computed 1-component inversions. This choice is based on the physical background of upward propagating slow magnetoacoustic waves in the umbral chromosphere. The magnetized solar atmosphere is characterized by a dimensionless parameter, i.e., plasma beta. This is the ratio of gas pressure to magnetic pressure, which is less than unity in the chromosphere (\citealt{2001SoPh..203...71G}). In this regime, the restoring force for the slow magnetoacoustic waves is dominated by the gas pressure over magnetic tension. These waves are essentially guided along the magnetic field lines, since across-field motion would require overcoming strong magnetic pressure. Therefore, along the field lines, the wave behaves like an acoustic wave, which is longitudinal in nature. Upward propagating acoustic waves are expected to not affect the magnetic field, thus not inducing magnetic field oscillations. The absence of magnetic field fluctuations associated with these waves is also supported by numerical modeling \citep[e.g.,][]{2021ApJ...918...47F}. We note that 2-component inversions allowing changes in the magnetic field can also reproduce the observed Stokes profiles, so our results cannot discard magnetic field fluctuations during shocks. Even if the analyzed wave modes are not associated with magnetic field fluctuations, opacity effects may still play a relevant role. However, we have restricted the inversion model to the minimum number of free parameters required to obtain a successful fitting of the profiles.

Our 2-component inversion setup for the shocked profiles uses a chromosphere with a positive velocity (downflow) on top of another chromosphere with a negative velocity (upflow). This mimics the expected arrival of the upward propagating shock wave. As the wavefront propagates upward, the downflowing plasma from higher layers falls on the upflowing plasma from below, forming the shock as an interphase with a sharp transition between flows with opposite directions. While the Stokes profiles generated from this configuration provide a good agreement with the observations, other configurations can also reproduce the observed profiles. \cite{2005ApJ...635..670C} found evidences of a two-component scenario in umbral oscillations observed in He I 10830 {\AA}, where two chromospheres added with a filling factor are present inside the resolution element. They were interpreted as an active oscillating component, alternating upflows and downflows, coexisting side by side with a quiet component nearly at rest. Their results support the model previously proposed by \cite{2000ApJ...544.1141S}. Our observations, acquired with higher spatial resolution, exhibit a different phenomenon. During the shocks, two chromospheres with opposite velocities must coexist inside the resolution element. Our favored scenario is that during the shocks, the He I 10830 {\AA} is sensitive to atmospheric layers at upstream and downstream parts of the wave, that is, at both sides of the shock front. Sharp velocity gradients up to 20 km s$^{-1}$ are required to fit the observed Stokes profiles (see Table \ref{2comp}).

\section{Conclusions}
\label{Sect5}
Our comprehensive analysis of the magnetoacoustic shocks and their impact on the spectropolarimetric signals from five different sunspots reveals the complex nature of the Stokes profiles from the shocked atmosphere. The results from one of our sunspots can potentially support the scenario of striking fluctuations in the response height of the He I 10830 {\AA} line \citep{2023A&A...676A..77F}, where higher values of the line-of-sight magnetic field accompanying the shock can be interpreted as a deeper atmospheric sampling of the He line. However, this scenario fails to explain the other four cases since they exhibit reduced magnetic fields during shocks (AR4 and AR5) or lack a significant correlation between the inferred field fluctuations and the magnetic field gradient with atmospheric height (AR1 and AR3). In the case of the latter, those two sunspots are further from disk center than AR2. The estimated magnetic field gradient from the photosphere to the chromosphere is influenced by projection effects, which might partially account for the weak correlations observed between magnetic field fluctuations and atmospheric gradients. Our analysis indicate that the magnetic field variations inferred from 1-component inversions may not reflect real changes in the field, but rather result from inverting the observed Stokes profiles with an inadequate model. A 2-component chromosphere model, where both chromospheres exhibit remarkably different Doppler velocities, can actually reproduce all the Stokes profiles associated with shocks. This result shows that, during some stages of the oscillation, the He I 10830 {\AA} line probes both sides of the shock fronts. The large velocity gradients associated with the shock are the cause of the observed complex Stokes profiles. Although the 2-component inversions do not require changes in the magnetic field to provide an excellent fit of the observed profiles, our study cannot discard magnetic field fluctuations produced by variations in the opacity of the He I 10830 {\AA} line taking place in atmospheres with vertical magnetic field gradients.


\begin{acknowledgments}
	
Financial support from grants PID2021-127487NB-I00, PID2024-156538NB-I00, and PID2024-156066OB-C55 funded by MICIU/AEI/ 10.13039/501100011033 and by ERDF, EU and from grant CNS2023-145233 funded by MICIU/AEI/10.13039/501100011033 and by “European Union NextGenerationEU/PRTR” is gratefully acknowledged. TF acknowledges grant RYC2020-030307-I funded by MCIN/AEI/ 10.13039/501100011033 and by “ESF Investing in your future”. CK acknowledges grant RYC2022-037660-I and SJGM grant RYC2022-037565-I, both funded by MCIN/AEI/10.13039/501100011033 and by ''ESF Investing in your future''. AAR acknowledges funding from the Agencia Estatal de Investigación del Ministerio de Ciencia, Innovación y Universidades (MCIU/AEI) under grant `Polarimetric Inference of Magnetic Fields' and the European Regional Development Fund (ERDF) with reference PID2022-136563NB-I00/10.13039/501100011033. The 1.5 m GREGOR solar telescope was built by a German consortium under the leadership of the Leibniz-Institut für Sonnenphysik in Freiburg in Freiburg with the Leibniz-Institut für Astrophysik Potsdam, the Institut für Astrophysik Göttingen, and the Max-Planck-Institut für Sonnen-systemforschung in Göttingen as partners, and with contributions by the Instituto de Astrofísica de Canarias and the Astronomical Institute of the Academy of Sciences of the Czech Republic. The redesign of the GREGOR AO and instrument `distribution optics was carried out by KIS whose technical staff is gratefully acknowledged.

\end{acknowledgments}

\bibliography{reference.bib}{}

@ARTICLE{1978AnSta...6..461S,
       author = {{Schwarz}, Gideon},
        title = "{Estimating the Dimension of a Model}",
      journal = {Annals of Statistics},
         year = 1978,
        month = jul,
       volume = {6},
       number = {2},
        pages = {461-464},
       adsurl = {https://ui.adsabs.harvard.edu/abs/1978AnSta...6..461S}
}

@ARTICLE{2012ApJ...748...83A,
       author = {{Asensio Ramos}, A. and {Manso Sainz}, R. and {Mart{\'\i}nez Gonz{\'a}lez}, M.~J. and {Viticchi{\'e}}, B. and {Orozco Su{\'a}rez}, D. and {Socas-Navarro}, H.},
        title = "{Model Selection for Spectropolarimetric Inversions}",
      journal = {\apj},
         year = 2012,
        month = apr,
       volume = {748},
       number = {2},
          eid = {83},
        pages = {83},
          doi = {10.1088/0004-637X/748/2/83},
archivePrefix = {arXiv},
       eprint = {1201.5063},
 primaryClass = {astro-ph.SR},
       adsurl = {https://ui.adsabs.harvard.edu/abs/2012ApJ...748...83A}
}

@ARTICLE{2012A&A...539A.131K,
       author = {{Kuckein}, C. and {Mart{\'\i}nez Pillet}, V. and {Centeno}, R.},
        title = "{An active region filament studied simultaneously in the chromosphere and photosphere. I. Magnetic structure}",
      journal = {\aap},
         year = 2012,
        month = mar,
       volume = {539},
          eid = {A131},
        pages = {A131},
          doi = {10.1051/0004-6361/201117675},
archivePrefix = {arXiv},
       eprint = {1112.1672},
 primaryClass = {astro-ph.SR},
       adsurl = {https://ui.adsabs.harvard.edu/abs/2012A&A...539A.131K}
}

@INPROCEEDINGS{2008SPIE.7019E..1EW,
       author = {{W{\"o}ger}, Friedrich and {von der L{\"u}he}, II, Oskar},
        title = "{KISIP: a software package for speckle interferometry of adaptive optics corrected solar data}",
    booktitle = {Advanced Software and Control for Astronomy II},
         year = 2008,
       editor = {{Bridger}, Alan and {Radziwill}, Nicole M.},
       series = {Society of Photo-Optical Instrumentation Engineers (SPIE) Conference Series},
       volume = {7019},
        month = jul,
          eid = {70191E},
        pages = {70191E},
          doi = {10.1117/12.788062},
       adsurl = {https://ui.adsabs.harvard.edu/abs/2008SPIE.7019E..1EW}
}

@INPROCEEDINGS{2017IAUS..327...20K,
       author = {{Kuckein}, C. and {Denker}, C. and {Verma}, M. and {Balthasar}, H. and {Gonz{\'a}lez Manrique}, S.~J. and {Louis}, R.~E. and {Diercke}, A.},
        title = "{sTools - a data reduction pipeline for the GREGOR Fabry-P{\'e}rot Interferometer and the High-resolution Fast Imager at the GREGOR solar telescope}",
    booktitle = {Fine Structure and Dynamics of the Solar Atmosphere},
         year = 2017,
       editor = {{Vargas Dom{\'\i}nguez}, S. and {Kosovichev}, A.~G. and {Antolin}, P. and {Harra}, L.},
       series = {IAU Symposium},
       volume = {327},
        month = oct,
        pages = {20-24},
          doi = {10.1017/S1743921317000114},
archivePrefix = {arXiv},
       eprint = {1701.01670},
 primaryClass = {astro-ph.IM},
       adsurl = {https://ui.adsabs.harvard.edu/abs/2017IAUS..327...20K}
}

@BOOK{2004ASSL..307.....L,
       author = {{Landi Degl'Innocenti}, E. and {Landolfi}, M.},
        title = "{Polarization in Spectral Lines}",
         year = 2004,
       volume = {307},
          doi = {10.1007/978-1-4020-2415-3},
       adsurl = {https://ui.adsabs.harvard.edu/abs/2004ASSL..307.....L}
}

@ARTICLE{2023JATIS...9a5001D,
       author = {{Denker}, Carsten and {Verma}, Meetu and {Wi{\'s}niewska}, Aneta and {Kamlah}, Robert and {Kontogiannis}, Ioannis and {Dineva}, Ekaterina and {Rendtel}, J{\"u}rgen and {Bauer}, Svend-Marian and {Dionies}, Mario and {{\"O}nel}, Hakan and {Woche}, Manfred and {Kuckein}, Christoph and {Seelemann}, Thomas and {Pal}, Partha S.},
        title = "{Improved High-resolution Fast Imager}",
      journal = {Journal of Astronomical Telescopes, Instruments, and Systems},
         year = 2023,
        month = jan,
       volume = {9},
          eid = {015001},
        pages = {015001},
          doi = {10.1117/1.JATIS.9.1.015001},
       adsurl = {https://ui.adsabs.harvard.edu/abs/2023JATIS...9a5001D}
}

@ARTICLE{2018ApJS..236....5D,
       author = {{Denker}, Carsten and {Kuckein}, Christoph and {Verma}, Meetu and {Gonz{\'a}lez Manrique}, Sergio J. and {Diercke}, Andrea and {Enke}, Harry and {Klar}, Jochen and {Balthasar}, Horst and {Louis}, Rohan E. and {Dineva}, Ekaterina},
        title = "{High-cadence Imaging and Imaging Spectroscopy at the GREGOR Solar Telescope{\textemdash}A Collaborative Research Environment for High-resolution Solar Physics}",
      journal = {\apjs},
         year = 2018,
        month = may,
       volume = {236},
       number = {1},
          eid = {5},
        pages = {5},
          doi = {10.3847/1538-4365/aab773},
archivePrefix = {arXiv},
       eprint = {1802.10146},
 primaryClass = {astro-ph.IM},
       adsurl = {https://ui.adsabs.harvard.edu/abs/2018ApJS..236....5D}
}

@ARTICLE{2005ApJ...635..670C,
       author = {{Centeno}, R. and {Socas-Navarro}, H. and {Collados}, M. and {Trujillo Bueno}, J.},
        title = "{Evidence for Fine Structure in the Chromospheric Umbral Oscillation}",
      journal = {\apj},
         year = 2005,
        month = dec,
       volume = {635},
       number = {1},
        pages = {670-673},
          doi = {10.1086/497393},
archivePrefix = {arXiv},
       eprint = {astro-ph/0510740},
 primaryClass = {astro-ph},
       adsurl = {https://ui.adsabs.harvard.edu/abs/2005ApJ...635..670C}
}

@ARTICLE{2024SoPh..299...23M,
       author = {{Modestov}, M. and {Khomenko}, E. and {Vitas}, N. and {de Vicente}, A. and {Navarro}, A. and {Gonz{\'a}lez-Morales}, P.~A. and {Collados}, M. and {Felipe}, T. and {Mart{\'\i}nez-G{\'o}mez}, D. and {Hunana}, P. and {Luna}, M. and {Koll Pistarini}, M. and {Popescu Braileanu}, B. and {Perdomo Garc{\'\i}a}, A. and {Liakh}, V. and {Santamaria}, I. and {Gomez Miguez}, M.~M.},
        title = "{MANCHA3D Code: Multipurpose Advanced Nonideal MHD Code for High-Resolution Simulations in Astrophysics}",
      journal = {\solphys},
         year = 2024,
        month = feb,
       volume = {299},
       number = {2},
          eid = {23},
        pages = {23},
          doi = {10.1007/s11207-024-02267-1},
archivePrefix = {arXiv},
       eprint = {2312.04179},
 primaryClass = {astro-ph.SR},
       adsurl = {https://ui.adsabs.harvard.edu/abs/2024SoPh..299...23M}
}

@ARTICLE{2024A&A...692A.169Q,
       author = {{Quintero Noda}, C. and {Shchukina}, N.~G. and {Asensio Ramos}, A. and {Mart{\'\i}nez Gonz{\'a}lez}, M.~J. and {del Pino Alem{\'a}n}, T. and {Trelles Arjona}, J.~C. and {Collados}, M.},
        title = "{Non-local thermodynamic equilibrium inversions of the Si I 10827 {\r{A}} spectral line}",
      journal = {\aap},
         year = 2024,
        month = dec,
       volume = {692},
          eid = {A169},
        pages = {A169},
          doi = {10.1051/0004-6361/202452233},
archivePrefix = {arXiv},
       eprint = {2411.12397},
 primaryClass = {astro-ph.SR},
       adsurl = {https://ui.adsabs.harvard.edu/abs/2024A&A...692A.169Q}
}

@ARTICLE{2019A&A...628A..47S,
       author = {{Shchukina}, N.~G. and {Trujillo Bueno}, J.},
        title = "{The diagnostic potential of the weak field approximation for investigating the quiet Sun magnetism: the Si I 10 827 {\r{A}} line}",
      journal = {\aap},
         year = 2019,
        month = aug,
       volume = {628},
          eid = {A47},
        pages = {A47},
          doi = {10.1051/0004-6361/201935510},
       adsurl = {https://ui.adsabs.harvard.edu/abs/2019A&A...628A..47S}
}

@ARTICLE{2017A&A...603A..98S,
       author = {{Shchukina}, N.~G. and {Sukhorukov}, A.~V. and {Trujillo Bueno}, J.},
        title = "{A Si I atomic model for NLTE spectropolarimetric diagnostics of the 10 827 {\r{A}} line}",
      journal = {\aap},
         year = 2017,
        month = jul,
       volume = {603},
          eid = {A98},
        pages = {A98},
          doi = {10.1051/0004-6361/201630236},
       adsurl = {https://ui.adsabs.harvard.edu/abs/2017A&A...603A..98S}
}

@ARTICLE{2012KPCB...28..169S,
       author = {{Sukhorukov}, A.~V. and {Shchukina}, N.~G.},
        title = "{NLTE formation of the solar silicon spectrum: Silicon abundance in one-dimensional models of the solar atmosphere}",
      journal = {Kinematics and Physics of Celestial Bodies},
         year = 2012,
        month = jul,
       volume = {28},
       number = {4},
        pages = {169-182},
          doi = {10.3103/S0884591312040071},
       adsurl = {https://ui.adsabs.harvard.edu/abs/2012KPCB...28..169S}
}

@ARTICLE{1994A&A...291..622C,
       author = {{Collados}, M. and {Martinez Pillet}, V. and {Ruiz Cobo}, B. and {del Toro Iniesta}, J.~C. and {Vazquez}, M.},
        title = "{Observed differences between large and small sunspots.}",
      journal = {\aap},
         year = 1994,
        month = nov,
       volume = {291},
        pages = {622-634},
       adsurl = {https://ui.adsabs.harvard.edu/abs/1994A&A...291..622C}
}

@ARTICLE{2020A&A...641A..27K,
       author = {{Kleint}, Lucia and {Berkefeld}, Thomas and {Esteves}, Miguel and {Sonner}, Thomas and {Volkmer}, Reiner and {Gerber}, Karin and {Kr{\"a}mer}, Felix and {Grassin}, Olivier and {Berdyugina}, Svetlana},
        title = "{GREGOR: Optics redesign and updates from 2018-2020}",
      journal = {\aap},
         year = 2020,
        month = sep,
       volume = {641},
          eid = {A27},
        pages = {A27},
          doi = {10.1051/0004-6361/202038208},
archivePrefix = {arXiv},
       eprint = {2006.11875},
 primaryClass = {astro-ph.IM},
       adsurl = {https://ui.adsabs.harvard.edu/abs/2020A&A...641A..27K}
}

@ARTICLE{2001SoPh..203...71G,
       author = {{Gary}, G. Allen},
        title = "{Plasma Beta above a Solar Active Region: Rethinking the Paradigm}",
      journal = {\solphys},
         year = 2001,
        month = oct,
       volume = {203},
       number = {1},
        pages = {71-86},
          doi = {10.1023/A:1012722021820},
       adsurl = {https://ui.adsabs.harvard.edu/abs/2001SoPh..203...71G}
}

@ARTICLE{2021ApJ...916....5S,
	author = {{Schad}, Thomas A. and {Dima}, Gabriel I. and {Anan}, Tetsu},
	title = "{He I Spectropolarimetry of a Supersonic Coronal Downflow Within a Sunspot Umbra}",
	journal = {\apj},
	year = 2021,
	month = jul,
	volume = {916},
	number = {1},
	eid = {5},
	pages = {5},
	doi = {10.3847/1538-4357/ac01eb},
	archivePrefix = {arXiv},
	eprint = {2105.12853},
	primaryClass = {astro-ph.SR},
	adsurl = {https://ui.adsabs.harvard.edu/abs/2021ApJ...916....5S}
}

@ARTICLE{2019A&A...621A..35L,
	author = {{Libbrecht}, Tine and {de la Cruz Rodr{\'\i}guez}, Jaime and {Danilovic}, Sanja and {Leenaarts}, Jorrit and {Pazira}, Hiva},
	title = "{Chromospheric condensations and magnetic field in a C3.6-class flare studied via He I D$_{3}$ spectro-polarimetry}",
	journal = {\aap},
	year = 2019,
	month = jan,
	volume = {621},
	eid = {A35},
	pages = {A35},
	doi = {10.1051/0004-6361/201833610},
	archivePrefix = {arXiv},
	eprint = {1806.06880},
	primaryClass = {astro-ph.SR},
	adsurl = {https://ui.adsabs.harvard.edu/abs/2019A&A...621A..35L}
}

@ARTICLE{2018PASJ...70..101A,
	author = {{Anan}, Tetsu and {Yoneya}, Takurou and {Ichimoto}, Kiyoshi and {UeNo}, Satoru and {Shiota}, Daikou and {Nozawa}, Satoshi and {Takasao}, Shinsuke and {Kawate}, Tomoko},
	title = "{Measurement of vector magnetic field in a flare kernel with a spectropolarimetric observation in He I 10830 {\r{A}}}",
	journal = {\pasj},
	year = 2018,
	month = dec,
	volume = {70},
	number = {6},
	eid = {101},
	pages = {101},
	doi = {10.1093/pasj/psy105},
	archivePrefix = {arXiv},
	eprint = {1808.06821},
	primaryClass = {astro-ph.SR},
	adsurl = {https://ui.adsabs.harvard.edu/abs/2018PASJ...70..101A}
}

@ARTICLE{1992ApJ...398..375R,
       author = {{Ruiz Cobo}, B. and {del Toro Iniesta}, J.~C.},
        title = "{Inversion of Stokes Profiles}",
      journal = {\apj},
         year = 1992,
        month = oct,
       volume = {398},
        pages = {375},
          doi = {10.1086/171862},
       adsurl = {https://ui.adsabs.harvard.edu/abs/1992ApJ...398..375R}
}

@ARTICLE{2008ApJ...683..542A,
       author = {{Asensio Ramos}, A. and {Trujillo Bueno}, J. and {Landi Degl'Innocenti}, E.},
        title = "{Advanced Forward Modeling and Inversion of Stokes Profiles Resulting from the Joint Action of the Hanle and Zeeman Effects}",
      journal = {\apj},
         year = 2008,
        month = aug,
       volume = {683},
       number = {1},
        pages = {542-565},
          doi = {10.1086/589433},
archivePrefix = {arXiv},
       eprint = {0804.2695},
 primaryClass = {astro-ph},
       adsurl = {https://ui.adsabs.harvard.edu/abs/2008ApJ...683..542A}
}

@INPROCEEDINGS{1999ASPC..184....3C,
       author = {{Collados}, M.},
        title = "{High Resolution Spectropolarimetry and Magnetography}",
    booktitle = {Third Advances in Solar Physics Euroconference: Magnetic Fields and Oscillations},
         year = 1999,
       editor = {{Schmieder}, B. and {Hofmann}, A. and {Staude}, J.},
       series = {Astronomical Society of the Pacific Conference Series},
       volume = {184},
        month = sep,
        pages = {3-22},
       adsurl = {https://ui.adsabs.harvard.edu/abs/1999ASPC..184....3C}
}

@INPROCEEDINGS{2003SPIE.4843...55C,
       author = {{Collados}, Manuel V.},
        title = "{Stokes polarimeters in the near-infrared}",
    booktitle = {Polarimetry in Astronomy},
         year = 2003,
       editor = {{Fineschi}, Silvano},
       series = {Society of Photo-Optical Instrumentation Engineers (SPIE) Conference Series},
       volume = {4843},
        month = feb,
        pages = {55-65},
          doi = {10.1117/12.459370},
       adsurl = {https://ui.adsabs.harvard.edu/abs/2003SPIE.4843...55C}
}

@ARTICLE{2012AN....333..796S,
       author = {{Schmidt}, W. and {von der L{\"u}he}, O. and {Volkmer}, R. and {Denker}, C. and {Solanki}, S.~K. and {Balthasar}, H. and {Bello Gonz{\'a}lez}, N. and {Berkefeld}, Th. and {Collados}, M. and {Fischer}, A. and {Halbgewachs}, C. and {Heidecke}, F. and {Hofmann}, A. and {Kneer}, F. and {Lagg}, A. and {Nicklas}, H. and {Popow}, E. and {Puschmann}, K.~G. and {Schmidt}, D. and {Sigwarth}, M. and {Sobotka}, M. and {Soltau}, D. and {Staude}, J. and {Strassmeier}, K.~G. and {Waldmann}, T.~A.},
        title = "{The 1.5 meter solar telescope GREGOR}",
      journal = {Astronomische Nachrichten},
         year = 2012,
        month = nov,
       volume = {333},
       number = {9},
        pages = {796-809},
          doi = {10.1002/asna.201211725},
       adsurl = {https://ui.adsabs.harvard.edu/abs/2012AN....333..796S}
}

@ARTICLE{2012AN....333..872C,
       author = {{Collados}, M. and {L{\'o}pez}, R. and {P{\'a}ez}, E. and {Hern{\'a}ndez}, E. and {Reyes}, M. and {Calcines}, A. and {Ballesteros}, E. and {D{\'\i}az}, J.~J. and {Denker}, C. and {Lagg}, A. and {Schlichenmaier}, R. and {Schmidt}, W. and {Solanki}, S.~K. and {Strassmeier}, K.~G. and {von der L{\"u}he}, O. and {Volkmer}, R.},
        title = "{GRIS: The GREGOR Infrared Spectrograph}",
      journal = {Astronomische Nachrichten},
         year = 2012,
        month = nov,
       volume = {333},
       number = {9},
        pages = {872},
          doi = {10.1002/asna.201211738},
       adsurl = {https://ui.adsabs.harvard.edu/abs/2012AN....333..872C}
}

@ARTICLE{2018ApJ...860...28H,
       author = {{Houston}, S.~J. and {Jess}, D.~B. and {Asensio Ramos}, A. and {Grant}, S.~D.~T. and {Beck}, C. and {Norton}, A.~A. and {Krishna Prasad}, S.},
        title = "{The Magnetic Response of the Solar Atmosphere to Umbral Flashes}",
      journal = {\apj},
         year = 2018,
        month = jun,
       volume = {860},
       number = {1},
          eid = {28},
        pages = {28},
          doi = {10.3847/1538-4357/aab366},
archivePrefix = {arXiv},
       eprint = {1803.00018},
 primaryClass = {astro-ph.SR},
       adsurl = {https://ui.adsabs.harvard.edu/abs/2018ApJ...860...28H}
}

@ARTICLE{2010ApJ...719..357F,
       author = {{Felipe}, T. and {Khomenko}, E. and {Collados}, M.},
        title = "{Magneto-acoustic Waves in Sunspots: First Results From a New Three-dimensional Nonlinear Magnetohydrodynamic Code}",
      journal = {\apj},
         year = 2010,
        month = aug,
       volume = {719},
       number = {1},
        pages = {357-377},
          doi = {10.1088/0004-637X/719/1/357},
archivePrefix = {arXiv},
       eprint = {1006.2998},
 primaryClass = {astro-ph.SR},
       adsurl = {https://ui.adsabs.harvard.edu/abs/2010ApJ...719..357F}
}

@ARTICLE{2006ApJ...653..739K,
       author = {{Khomenko}, E. and {Collados}, M.},
        title = "{Numerical Modeling of Magnetohydrodynamic Wave Propagation and Refraction in Sunspots}",
      journal = {\apj},
         year = 2006,
        month = dec,
       volume = {653},
       number = {1},
        pages = {739-755},
          doi = {10.1086/507760},
       adsurl = {https://ui.adsabs.harvard.edu/abs/2006ApJ...653..739K}
}

@ARTICLE{2021ApJ...918...47F,
       author = {{Felipe}, Tobias and {Socas Navarro}, Hector and {Sangeetha}, C.~R. and {Milic}, Ivan},
        title = "{Limitations of the Ca II 8542 {\r{A}} Line for the Determination of Magnetic Field Oscillations}",
      journal = {\apj},
         year = 2021,
        month = sep,
       volume = {918},
       number = {2},
          eid = {47},
        pages = {47},
          doi = {10.3847/1538-4357/ac111c},
archivePrefix = {arXiv},
       eprint = {2107.02160},
 primaryClass = {astro-ph.SR},
       adsurl = {https://ui.adsabs.harvard.edu/abs/2021ApJ...918...47F}
}

@ARTICLE{2015A&A...577A...7S,
       author = {{Socas-Navarro}, H. and {de la Cruz Rodr{\'\i}guez}, J. and {Asensio Ramos}, A. and {Trujillo Bueno}, J. and {Ruiz Cobo}, B.},
        title = "{An open-source, massively parallel code for non-LTE synthesis and inversion of spectral lines and Zeeman-induced Stokes profiles}",
      journal = {\aap},
         year = 2015,
        month = may,
       volume = {577},
          eid = {A7},
        pages = {A7},
          doi = {10.1051/0004-6361/201424860},
archivePrefix = {arXiv},
       eprint = {1408.6101},
 primaryClass = {astro-ph.SR},
       adsurl = {https://ui.adsabs.harvard.edu/abs/2015A&A...577A...7S}
}

@ARTICLE{2017ApJ...845..102H,
       author = {{Henriques}, V.~M.~J. and {Mathioudakis}, M. and {Socas-Navarro}, H. and {de la Cruz Rodr{\'\i}guez}, J.},
        title = "{A Hot Downflowing Model Atmosphere for Umbral Flashes and the Physical Properties of Their Dark Fibrils}",
      journal = {\apj},
         year = 2017,
        month = aug,
       volume = {845},
       number = {2},
          eid = {102},
        pages = {102},
          doi = {10.3847/1538-4357/aa7ca4},
archivePrefix = {arXiv},
       eprint = {1706.05311},
 primaryClass = {astro-ph.SR},
       adsurl = {https://ui.adsabs.harvard.edu/abs/2017ApJ...845..102H}
}

@ARTICLE{2007ApJ...671.1005B,
       author = {{Bloomfield}, D. Shaun and {Lagg}, Andreas and {Solanki}, Sami K.},
        title = "{The Nature of Running Penumbral Waves Revealed}",
      journal = {\apj},
         year = 2007,
        month = dec,
       volume = {671},
       number = {1},
        pages = {1005-1012},
          doi = {10.1086/523266},
archivePrefix = {arXiv},
       eprint = {0709.3731},
 primaryClass = {astro-ph},
       adsurl = {https://ui.adsabs.harvard.edu/abs/2007ApJ...671.1005B}
}

@ARTICLE{2023ApJ...945L..27F,
       author = {{French}, Ryan J. and {Bogdan}, Thomas J. and {Casini}, Roberto and {de Wijn}, Alfred G. and {Judge}, Philip G.},
        title = "{First Observation of Chromospheric Waves in a Sunspot by DKIST/ViSP: The Anatomy of an Umbral Flash}",
      journal = {\apjl},
         year = 2023,
        month = mar,
       volume = {945},
       number = {2},
          eid = {L27},
        pages = {L27},
          doi = {10.3847/2041-8213/acb8b5},
archivePrefix = {arXiv},
       eprint = {2303.06105},
 primaryClass = {astro-ph.SR},
       adsurl = {https://ui.adsabs.harvard.edu/abs/2023ApJ...945L..27F}
}

@ARTICLE{2009ApJ...696.1683S,
       author = {{Socas-Navarro}, H. and {McIntosh}, S.~W. and {Centeno}, R. and {de Wijn}, A.~G. and {Lites}, B.~W.},
        title = "{Direct Imaging of Fine Structure in the Chromosphere of a Sunspot Umbra}",
      journal = {\apj},
         year = 2009,
        month = may,
       volume = {696},
       number = {2},
        pages = {1683-1688},
          doi = {10.1088/0004-637X/696/2/1683},
archivePrefix = {arXiv},
       eprint = {0810.0597},
 primaryClass = {astro-ph},
       adsurl = {https://ui.adsabs.harvard.edu/abs/2009ApJ...696.1683S}
}

@ARTICLE{2019ApJ...882..161A,
       author = {{Anan}, Tetsu and {Schad}, Thomas A. and {Jaeggli}, Sarah A. and {Tarr}, Lucas A.},
        title = "{Shock Heating Energy of Umbral Flashes Measured with Integral Field Unit Spectroscopy}",
      journal = {\apj},
         year = 2019,
        month = sep,
       volume = {882},
       number = {2},
          eid = {161},
        pages = {161},
          doi = {10.3847/1538-4357/ab357f},
archivePrefix = {arXiv},
       eprint = {1907.10797},
 primaryClass = {astro-ph.SR},
       adsurl = {https://ui.adsabs.harvard.edu/abs/2019ApJ...882..161A}
}

@ARTICLE{2020ApJ...892...49H,
       author = {{Houston}, S.~J. and {Jess}, D.~B. and {Keppens}, R. and {Stangalini}, M. and {Keys}, P.~H. and {Grant}, S.~D.~T. and {Jafarzadeh}, S. and {McFetridge}, L.~M. and {Murabito}, M. and {Ermolli}, I. and {Giorgi}, F.},
        title = "{Magnetohydrodynamic Nonlinearities in Sunspot Atmospheres: Chromospheric Detections of Intermediate Shocks}",
      journal = {\apj},
         year = 2020,
        month = mar,
       volume = {892},
       number = {1},
          eid = {49},
        pages = {49},
          doi = {10.3847/1538-4357/ab7a90},
archivePrefix = {arXiv},
       eprint = {2002.12368},
 primaryClass = {astro-ph.SR},
       adsurl = {https://ui.adsabs.harvard.edu/abs/2020ApJ...892...49H}
}

@ARTICLE{2020A&A...642A.215H,
       author = {{Henriques}, Vasco M.~J. and {Nelson}, Chris J. and {Rouppe van der Voort}, Luc H.~M. and {Mathioudakis}, Mihalis},
        title = "{Umbral chromospheric fine structure and umbral flashes modelled as one: The corrugated umbra}",
      journal = {\aap},
         year = 2020,
        month = oct,
       volume = {642},
          eid = {A215},
        pages = {A215},
          doi = {10.1051/0004-6361/202038538},
archivePrefix = {arXiv},
       eprint = {2008.05482},
 primaryClass = {astro-ph.SR},
       adsurl = {https://ui.adsabs.harvard.edu/abs/2020A&A...642A.215H}
}

@ARTICLE{2018A&A...619A..63J,
       author = {{Joshi}, Jayant and {de la Cruz Rodr{\'\i}guez}, Jaime},
        title = "{Magnetic field variations associated with umbral flashes and penumbral waves}",
      journal = {\aap},
         year = 2018,
        month = nov,
       volume = {619},
          eid = {A63},
        pages = {A63},
          doi = {10.1051/0004-6361/201832955},
archivePrefix = {arXiv},
       eprint = {1803.01737},
 primaryClass = {astro-ph.SR},
       adsurl = {https://ui.adsabs.harvard.edu/abs/2018A&A...619A..63J}
}

@ARTICLE{2018A&A...614A..73F,
       author = {{Felipe}, T. and {Socas-Navarro}, H. and {Przybylski}, D.},
        title = "{Inversions of synthetic umbral flashes: Effects of scanning time on the inferred atmospheres}",
      journal = {\aap},
         year = 2018,
        month = jun,
       volume = {614},
          eid = {A73},
        pages = {A73},
          doi = {10.1051/0004-6361/201732169},
archivePrefix = {arXiv},
       eprint = {1802.05028},
 primaryClass = {astro-ph.SR},
       adsurl = {https://ui.adsabs.harvard.edu/abs/2018A&A...614A..73F}
}

@ARTICLE{2019A&A...632A..75F,
       author = {{Felipe}, T. and {Esteban Pozuelo}, S.},
        title = "{Inversions of synthetic umbral flashes: a selection of wavelength sampling}",
      journal = {\aap},
         year = 2019,
        month = dec,
       volume = {632},
          eid = {A75},
        pages = {A75},
          doi = {10.1051/0004-6361/201936679},
archivePrefix = {arXiv},
       eprint = {1910.13980},
 primaryClass = {astro-ph.SR},
       adsurl = {https://ui.adsabs.harvard.edu/abs/2019A&A...632A..75F}
}

@ARTICLE{2021A&A...645L..12F,
       author = {{Felipe}, T. and {Henriques}, V.~M.~J. and {de la Cruz Rodr{\'\i}guez}, J. and {Socas-Navarro}, H.},
        title = "{Downflowing umbral flashes as evidence of standing waves in sunspot umbrae}",
      journal = {\aap},
         year = 2021,
        month = jan,
       volume = {645},
          eid = {L12},
        pages = {L12},
          doi = {10.1051/0004-6361/202039966},
archivePrefix = {arXiv},
       eprint = {2101.04188},
 primaryClass = {astro-ph.SR},
       adsurl = {https://ui.adsabs.harvard.edu/abs/2021A&A...645L..12F}
}

@ARTICLE{2020ApJ...896..150Y,
       author = {{Yurchyshyn}, Vasyl and {Kilcik}, Ali and {{\c{S}}ahin}, Seray and {Abramenko}, Valentina and {Lim}, Eun-Kyung},
        title = "{Spatial Distribution of the Origin of Umbral Waves in a Sunspot Umbra}",
      journal = {\apj},
         year = 2020,
        month = jun,
       volume = {896},
       number = {2},
          eid = {150},
        pages = {150},
          doi = {10.3847/1538-4357/ab91b8},
archivePrefix = {arXiv},
       eprint = {2005.04202},
 primaryClass = {astro-ph.SR},
       adsurl = {https://ui.adsabs.harvard.edu/abs/2020ApJ...896..150Y}
}

@ARTICLE{2015ApJ...800..129M,
       author = {{Madsen}, Chad A. and {Tian}, Hui and {DeLuca}, Edward E.},
        title = "{Observations of Umbral Flashes and Running Sunspot Waves with the Interface Region Imaging Spectrograph}",
      journal = {\apj},
         year = 2015,
        month = feb,
       volume = {800},
       number = {2},
          eid = {129},
        pages = {129},
          doi = {10.1088/0004-637X/800/2/129},
       adsurl = {https://ui.adsabs.harvard.edu/abs/2015ApJ...800..129M}
}

@ARTICLE{2025A&A...693A.165F,
       author = {{Felipe}, T. and {Gonz{\'a}lez Manrique}, S.~J. and {Mart{\'\i}nez-G{\'o}mez}, D. and {G{\'o}mez-M{\'\i}guez}, M.~M. and {Khomenko}, E. and {Quintero Noda}, C. and {Socas-Navarro}, H.},
        title = "{Observations of umbral flashes in the resonant sunspot chromosphere}",
      journal = {\aap},
         year = 2025,
        month = jan,
       volume = {693},
          eid = {A165},
        pages = {A165},
          doi = {10.1051/0004-6361/202452317},
archivePrefix = {arXiv},
       eprint = {2411.16467},
 primaryClass = {astro-ph.SR},
       adsurl = {https://ui.adsabs.harvard.edu/abs/2025A&A...693A.165F}
}

@ARTICLE{2000Sci...288.1396S,
       author = {{Socas-Navarro}, H. and {Trujillo Bueno}, J. and {Ruiz Cobo}, B.},
        title = "{Anomalous Polarization Profiles in Sunspots: Possible Origin of Umbral Flashes}",
      journal = {Science},
         year = 2000,
        month = may,
       volume = {288},
       number = {5470},
        pages = {1396-1398},
          doi = {10.1126/science.288.5470.1396},
       adsurl = {https://ui.adsabs.harvard.edu/abs/2000Sci...288.1396S}
}

@ARTICLE{2000ApJ...544.1141S,
       author = {{Socas-Navarro}, H. and {Trujillo Bueno}, J. and {Ruiz Cobo}, B.},
        title = "{Anomalous Circular Polarization Profiles in Sunspot Chromospheres}",
      journal = {\apj},
         year = 2000,
        month = dec,
       volume = {544},
       number = {2},
        pages = {1141-1154},
          doi = {10.1086/317261},
       adsurl = {https://ui.adsabs.harvard.edu/abs/2000ApJ...544.1141S}
}

@ARTICLE{2013A&A...556A.115D,
       author = {{de la Cruz Rodr{\'\i}guez}, J. and {Rouppe van der Voort}, L. and {Socas-Navarro}, H. and {van Noort}, M.},
        title = "{Physical properties of a sunspot chromosphere with umbral flashes}",
      journal = {\aap},
         year = 2013,
        month = aug,
       volume = {556},
          eid = {A115},
        pages = {A115},
          doi = {10.1051/0004-6361/201321629},
archivePrefix = {arXiv},
       eprint = {1304.0752},
 primaryClass = {astro-ph.SR},
       adsurl = {https://ui.adsabs.harvard.edu/abs/2013A&A...556A.115D}
}

@ARTICLE{2022ApJ...927..201A,
       author = {{Albidah}, A.~B. and {Fedun}, V. and {Aldhafeeri}, A.~A. and {Ballai}, I. and {Brevis}, W. and {Jess}, D.~B. and {Higham}, J. and {Stangalini}, M. and {Silva}, S.~S.~A. and {Verth}, G.},
        title = "{Magnetohydrodynamic Wave Mode Identification in Circular and Elliptical Sunspot Umbrae: Evidence for High-order Modes}",
      journal = {\apj},
         year = 2022,
        month = mar,
       volume = {927},
       number = {2},
          eid = {201},
        pages = {201},
          doi = {10.3847/1538-4357/ac51d9},
archivePrefix = {arXiv},
       eprint = {2202.00624},
 primaryClass = {astro-ph.SR},
       adsurl = {https://ui.adsabs.harvard.edu/abs/2022ApJ...927..201A}
}

@ARTICLE{2021RSPTA.37900172G,
       author = {{Gilchrist-Millar}, Caitlin A. and {Jess}, David B. and {Grant}, Samuel D.~T. and {Keys}, Peter H. and {Beck}, Christian and {Jafarzadeh}, Shahin and {Riedl}, Julia M. and {Van Doorsselaere}, Tom and {Ruiz Cobo}, Basilio},
        title = "{Magnetoacoustic wave energy dissipation in the atmosphere of solar pores}",
      journal = {Philosophical Transactions of the Royal Society of London Series A},
         year = 2021,
        month = feb,
       volume = {379},
       number = {2190},
          eid = {20200172},
        pages = {20200172},
          doi = {10.1098/rsta.2020.0172},
archivePrefix = {arXiv},
       eprint = {2007.11594},
 primaryClass = {astro-ph.SR},
       adsurl = {https://ui.adsabs.harvard.edu/abs/2021RSPTA.37900172G}
}

@ARTICLE{2017ApJ...842...59J,
       author = {{Jess}, David B. and {Van Doorsselaere}, Tom and {Verth}, Gary and {Fedun}, Viktor and {Krishna Prasad}, S. and {Erd{\'e}lyi}, Robert and {Keys}, Peter H. and {Grant}, Samuel D.~T. and {Uitenbroek}, Han and {Christian}, Damian J.},
        title = "{An Inside Look at Sunspot Oscillations with Higher Azimuthal Wavenumbers}",
      journal = {\apj},
         year = 2017,
        month = jun,
       volume = {842},
       number = {1},
          eid = {59},
        pages = {59},
          doi = {10.3847/1538-4357/aa73d6},
archivePrefix = {arXiv},
       eprint = {1705.06282},
 primaryClass = {astro-ph.SR},
       adsurl = {https://ui.adsabs.harvard.edu/abs/2017ApJ...842...59J}
}

@ARTICLE{2015ApJ...812L..15K,
       author = {{Krishna Prasad}, S. and {Jess}, D.~B. and {Khomenko}, Elena},
        title = "{On the Source of Propagating Slow Magnetoacoustic Waves in Sunspots}",
      journal = {\apjl},
         year = 2015,
        month = oct,
       volume = {812},
       number = {1},
          eid = {L15},
        pages = {L15},
          doi = {10.1088/2041-8205/812/1/L15},
archivePrefix = {arXiv},
       eprint = {1510.03275},
 primaryClass = {astro-ph.SR},
       adsurl = {https://ui.adsabs.harvard.edu/abs/2015ApJ...812L..15K}
}

@ARTICLE{2015LRSP...12....6K,
       author = {{Khomenko}, Elena and {Collados}, Manuel},
        title = "{Oscillations and Waves in Sunspots}",
      journal = {Living Reviews in Solar Physics},
         year = 2015,
        month = dec,
       volume = {12},
       number = {1},
          eid = {6},
        pages = {6},
          doi = {10.1007/lrsp-2015-6},
       adsurl = {https://ui.adsabs.harvard.edu/abs/2015LRSP...12....6K}
}

@ARTICLE{2006ApJ...640.1153C,
       author = {{Centeno}, Rebecca and {Collados}, Manuel and {Trujillo Bueno}, Javier},
        title = "{Spectropolarimetric Investigation of the Propagation of Magnetoacoustic Waves and Shock Formation in Sunspot Atmospheres}",
      journal = {\apj},
         year = 2006,
        month = apr,
       volume = {640},
       number = {2},
        pages = {1153-1162},
          doi = {10.1086/500185},
archivePrefix = {arXiv},
       eprint = {astro-ph/0512096},
 primaryClass = {astro-ph},
       adsurl = {https://ui.adsabs.harvard.edu/abs/2006ApJ...640.1153C}
}

@ARTICLE{2023JASTP.24706071K,
       author = {{Kumar}, Hirdesh and {Kumar}, Brajesh and {Rajaguru}, S.~P. and {Mathew}, Shibu K. and {Bayanna}, Ankala Raja},
        title = "{A study of the propagation of magnetoacoustic waves in small-scale magnetic fields using solar photospheric and chromospheric Dopplergrams: HMI/SDO and MAST observations}",
      journal = {Journal of Atmospheric and Solar-Terrestrial Physics},
         year = 2023,
        month = jun,
       volume = {247},
          eid = {106071},
        pages = {106071},
          doi = {10.1016/j.jastp.2023.106071},
archivePrefix = {arXiv},
       eprint = {2304.13492},
 primaryClass = {astro-ph.SR},
       adsurl = {https://ui.adsabs.harvard.edu/abs/2023JASTP.24706071K}
}

@ARTICLE{2019ApJ...871..155R,
       author = {{Rajaguru}, S.~P. and {Sangeetha}, C.~R. and {Tripathi}, Durgesh},
        title = "{Magnetic Fields and the Supply of Low-frequency Acoustic Wave Energy to the Solar Chromosphere}",
      journal = {\apj},
         year = 2019,
        month = feb,
       volume = {871},
       number = {2},
          eid = {155},
        pages = {155},
          doi = {10.3847/1538-4357/aaf883},
archivePrefix = {arXiv},
       eprint = {1812.05322},
 primaryClass = {astro-ph.SR},
       adsurl = {https://ui.adsabs.harvard.edu/abs/2019ApJ...871..155R}
}

@ARTICLE{2011ApJ...735...65F,
       author = {{Felipe}, T. and {Khomenko}, E. and {Collados}, M.},
        title = "{Magnetoacoustic Wave Energy from Numerical Simulations of an Observed Sunspot Umbra}",
      journal = {\apj},
         year = 2011,
        month = jul,
       volume = {735},
       number = {1},
          eid = {65},
        pages = {65},
          doi = {10.1088/0004-637X/735/1/65},
archivePrefix = {arXiv},
       eprint = {1104.4138},
 primaryClass = {astro-ph.SR},
       adsurl = {https://ui.adsabs.harvard.edu/abs/2011ApJ...735...65F}
}

@ARTICLE{2010ApJ...722..131F,
       author = {{Felipe}, T. and {Khomenko}, E. and {Collados}, M. and {Beck}, C.},
        title = "{Multi-layer Study of Wave Propagation in Sunspots}",
      journal = {\apj},
         year = 2010,
        month = oct,
       volume = {722},
       number = {1},
        pages = {131-144},
          doi = {10.1088/0004-637X/722/1/131},
archivePrefix = {arXiv},
       eprint = {1008.4004},
 primaryClass = {astro-ph.SR},
       adsurl = {https://ui.adsabs.harvard.edu/abs/2010ApJ...722..131F}
}

@ARTICLE{2016ApJ...830L..17Z,
       author = {{Zhao}, Junwei and {Felipe}, Tob{\'\i}as and {Chen}, Ruizhu and {Khomenko}, Elena},
        title = "{Tracing p-mode Waves from the Photosphere to the Corona in Active Regions}",
      journal = {\apjl},
         year = 2016,
        month = oct,
       volume = {830},
       number = {1},
          eid = {L17},
        pages = {L17},
          doi = {10.3847/2041-8205/830/1/L17},
       adsurl = {https://ui.adsabs.harvard.edu/abs/2016ApJ...830L..17Z}
}

@ARTICLE{2019A&A...621A..43F,
       author = {{Felipe}, T. and {Kuckein}, C. and {Khomenko}, E. and {Thaler}, I.},
        title = "{Spiral-shaped wavefronts in a sunspot umbra}",
      journal = {\aap},
         year = 2019,
        month = jan,
       volume = {621},
          eid = {A43},
        pages = {A43},
          doi = {10.1051/0004-6361/201834367},
archivePrefix = {arXiv},
       eprint = {1810.11257},
 primaryClass = {astro-ph.SR},
       adsurl = {https://ui.adsabs.harvard.edu/abs/2019A&A...621A..43F}
}

@ARTICLE{2023A&A...676A..77F,
       author = {{Felipe}, T. and {Gonz{\'a}lez Manrique}, S.~J. and {Sangeetha}, C.~R. and {Asensio Ramos}, A.},
        title = "{Magnetic field fluctuations in the shocked umbral chromosphere}",
      journal = {\aap},
         year = 2023,
        month = aug,
       volume = {676},
          eid = {A77},
        pages = {A77},
          doi = {10.1051/0004-6361/202244519},
archivePrefix = {arXiv},
       eprint = {2307.01313},
 primaryClass = {astro-ph.SR},
       adsurl = {https://ui.adsabs.harvard.edu/abs/2023A&A...676A..77F}
}

@incollection{Goedbloed,
  author    = {J.P.H. Goedbloed, S. Poedts},
  year      = {2004},
  title     = { },
  editor    = { },
  booktitle = {Principles of Magnetohydrodynamics},
  publisher = {Cambridge University Press},
  address   = {Cambridge},
  pages     = { }
}

@ARTICLE{1969SoPh....7..351B,
       author = {{Beckers}, Jacques M. and {Tallant}, Paul E.},
        title = "{Chromospheric Inhomogeneities in Sunspot Umbrae}",
      journal = {\solphys},
         year = 1969,
        month = jun,
       volume = {7},
       number = {3},
        pages = {351-365},
          doi = {10.1007/BF00146140},
       adsurl = {https://ui.adsabs.harvard.edu/abs/1969SoPh....7..351B}
}

@ARTICLE{1969SoPh....7..366W,
       author = {{Wittmann}, A.},
        title = "{Some Properties of Umbral Flashes}",
      journal = {\solphys},
         year = 1969,
        month = jun,
       volume = {7},
       number = {3},
        pages = {366-369},
          doi = {10.1007/BF00146141},
       adsurl = {https://ui.adsabs.harvard.edu/abs/1969SoPh....7..366W}
}
\bibliographystyle{aasjournal}



\end{document}